\documentclass[aps,pra,floatfix,amsmath,amssymb,eqsecnum,nofootinbib,twocolumn,superscriptaddress]{revtex4-2}

\usepackage{subcaption}
\usepackage{ragged2e}
\usepackage{graphicx}
\usepackage{algorithm}
\usepackage{hyperref}
\hypersetup{colorlinks,allcolors=black,breaklinks=true}
\usepackage{algpseudocodex}
\usepackage{tikz}[wire types={q, n}]
\usetikzlibrary{quantikz2}
\usetikzlibrary{fit, shapes, positioning}

\definecolor{allocation_fill}{HTML}{F5CED7}
\definecolor{allocation_border}{HTML}{F43764}
\definecolor{neutral_fill}{HTML}{98EBEF}
\definecolor{neutral_border}{HTML}{119DA4}
\definecolor{deallocation_fill}{HTML}{E5F79C}
\definecolor{deallocation_border}{HTML}{92AC2D}

\begin{document}

\title{Scalable Memory Recycling for Large Quantum Programs}

\author{Israel Reichental}
\affiliation{Classiq Technologies}

\author{Ravid Alon}
\affiliation{Classiq Technologies}

\author{Lior Preminger}
\affiliation{Classiq Technologies}

\author{Matan Vax}
\affiliation{Classiq Technologies}

\author{Amir Naveh}
\affiliation{Classiq Technologies}

\date{\today}

\begin{abstract}
    As quantum computing technology advances, the complexity of quantum algorithms increases, necessitating a shift from low-level circuit descriptions to high-level programming paradigms. This paper addresses the challenges of developing a compilation algorithm that optimizes memory management and scales well for bigger, more complex circuits. Our approach models the high-level quantum code as a control flow graph and presents a workflow that searches for a topological sort that maximizes opportunities for qubit reuse. Various heuristics for qubit reuse strategies handle the trade-off between circuit width and depth. We also explore scalability issues in large circuits, suggesting methods to mitigate compilation bottlenecks. By analyzing the structure of the circuit, we are able to identify sub-problems that can be solved separately, without a significant effect on circuit quality, while reducing runtime significantly. This method lays the groundwork for future advancements in quantum programming and compiler optimization by incorporating scalability into quantum memory management.
\end{abstract}

\maketitle

\section{Introduction}

As quantum hardware evolves to support the execution of larger quantum programs, the complexity of quantum algorithms scales up accordingly. To facilitate the design of such algorithms, the quantum software paradigm needs to shift from a circuit-level description, where the developer works with gates and qubits, into a higher-level approach that centers around functions and variables. A high-level description allows the quantum programmer to focus on the functional intent of the code, while an automated compilation pipeline hides the implementation details, such as memory management, runtime reduction, and more. Given the current scarcity of computational resources in quantum computing, it is crucial that quantum compilers provide optimal quantum executables.

One of the key challenges in designing quantum compilers is the inherent reversibility requirement of quantum computation. Unlike their classical analog, temporary quantum variables that store intermediate calculation results cannot be discarded outright. Instead, they should be \textit{uncomputed} \cite{5391327, Nielsen_Chuang_2010} by applying the conjugated operation after their values are no longer required for the computation. A sufficient condition for the conjugation to properly uncompute the variable is that it underwent classical operations (e.g., arithmetic) and the inputs for these operations haven’t been modified during the computation \cite{Rand_2019}.

The added uncomputation operations increase the circuit size, and optimizing the resulting resource demands is an issue undergoing extensive research \cite{doi:10.1137/0218053, meuli19, paradis21, paradis24, 10.1109/ISCA45697.2020.00054}. A crucial aspect of this optimization is the compiler's ability to efficiently reuse those qubits freed during uncomputation for subsequent operations. Without reuse, the qubit consumption can rise dramatically, rendering the execution of the quantum program infeasible due to excess resource demands.

We present a workflow to address this problem by transforming high-level quantum code into a control flow graph and executing a series of compilation passes on this graph to obtain a scheme for scheduling the uncomputation operations and reusing the freed qubits. Our approach allows for customization of the qubit reuse scheme according to user preferences. For example, they can optimize the circuit to use fewer qubits or allow quantum code segments to run in parallel to reduce execution runtime.

This paper is structured as follows. In Section \ref{sec:control-flow-graph}, we begin by discussing high-level quantum programming and variable semantics. This produces an intermediate representation that captures the quantum data flow and enables analysis of the dependency relations between the quantum operations. In Section \ref{sec:topological-sorting}, we then present an analysis pass that topologically sorts the control flow graph to facilitate the recycling of qubits released by the uncomputation operations. In Section \ref{sec:reuse-analysis-pass}, we examine various qubit reuse options based on this sort and analyze the trade-offs between qubit minimization and execution runtime. Next, in Section \ref{sec:scalability} we discuss the scalability of this compilation stage and outline measures to prevent it from becoming a compilation bottleneck in very large circuits. Finally, we discuss related work in Section \ref{sec:related-work}, and consider possible extensions to our work in Section \ref{sec:future-work}. Section \ref{sec:summary} is the Summary.

This work has been integrated into Classiq's platform for quantum software development \cite{classiq}.

\section{Control Flow Graph}\label{sec:control-flow-graph}

Most quantum programming representations allow users to express quantum operations by instantiating a gate and variables that the gate consumes\cite{openqasm, qir}. Here is a simple "Hello World" pseudo-code example that generates the first bell state. A quantum variable $qarr$ is declared and initialized to the zero state, and then Hadamard and CX gates act upon it.

\begin{algorithm}[H]
\begin{algorithmic}[1]
    \State $qarr \gets |00\rangle$
    \State $H\left(qarr\left[0\right]\right)$
    \State $CX\left(qarr\left[0\right], qarr\left[1\right]\right)$
\end{algorithmic}
\end{algorithm}

Higher-level quantum programming languages \cite{bichsel20, vax2025qmodexpressivehighlevelquantum, qiskit, qsharp, qrisp, tket, cuda, pennylane, cirq, tower} also support more complex concepts. The most common example is reusable code parts such as functions and closures. More advanced constructs range from quantum “if” statements that correspond to quantum control operations, to creating local variables and declaring them 'freed' or uncomputed so their qubits can be recycled. Here is a pseudo-code example implementing a generic control statement on a single operation. It evaluates the condition into a temporary one-qubit variable and applies it as a canonical control on the operation.

\begin{algorithm}[H]
\begin{algorithmic}[1]
    \Function{CONTROL}{$condition$, $operation$}
        \State Initialize: $aux \gets |0\rangle$
        \State $aux \gets$ Evaluate $condition$
        \If{$aux == |1\rangle$}
            \State $operation()$
        \EndIf
        \State Uncompute: $aux \gets |0\rangle$
    \EndFunction
\end{algorithmic}
\end{algorithm}

We can execute the function under different conditions and operations:

\begin{algorithm}[H]
\begin{algorithmic}[1]
\State $CONTROL(condition_1, operation_1)$
\State $\vdots$
\State $CONTROL(condition_k, operation_k)$
\end{algorithmic}
\end{algorithm}

Each loop iteration draws a qubit from the pool and then returns it, allowing these qubits to be reused between different calls. Having the compiler apply such a reuse scheme requires dependency relations between different operations. An operation B that follows another operation A can reuse its qubits. If two operations, A and B, consume disjoint sets of quantum variables, then A can reuse the qubits of B, B can reuse the qubits of A, or we can skip the reuse operation and run these operations in parallel to reduce the circuit depth.

To establish these relations, we form the operation control flow graph by translating the high-level description of variables into an intermediate representation that illustrates the flow of quantum data between the operations. The intermediate representation then translates into the control flow graph. The control flow graph resembles a quantum circuit, but it abstracts away the individual qubits associated with the quantum variables, implementation details of scheduling, the drawing of qubits from the pool, and the use of auxiliary qubits. The result is a directed acyclic graph (DAG), where each node is a quantum operation. Operations can be either atomic, such as native gates or qubit allocation, or compound, containing a sequence of atomic operations. We address this difference in detail in Section \ref{sec:scalability}, which discusses scalability. The edges in the graph represent the flow. An edge is added between two operations that are applied consecutively to the same variable.

We treat each node in the graph as a functional black box that has one of the following roles:

\begin{enumerate}
    \item Qubit-requiring node or an allocation node - A node that draws qubits from the pool of qubits in the $|0\rangle$ state. The qubits can be previously released qubits or new qubits drawn from the device pool (i.e., a quantum computer or a simulator).
    \item Qubit-releasing node or a deallocation node - A node that releases qubits to the pool.
    \item Neutral node - A node that is neither.
\end{enumerate}

We impose a restriction such that a node cannot simultaneously draw and release non-overlapping qubits. Although such nodes can be created artificially by arbitrarily combining allocation and deallocation nodes into a single node, our practical experience dictates that we can exclude these scenarios from our discussion without worrying about possible performance implications.

The problem description includes the total number of qubits required (type 1) or released (type 2) per node.

Note that qubit allocation or deallocation is not the mere functionality of nodes marked as such. They can also include Input/Output (I/O) functional qubits. To clarify this by an example, consider an allocation node in the form of a quantum out-of-place adder: its arguments are I/Os that remain unchanged, while the result variable uses qubits drawn from the pool. Similarly, the uncomputation of the addition result forms a deallocation node.

Each node can also incorporate \textit{auxiliary qubits}, which are both allocated and released by the node. Auxiliary qubits can be a part of the function’s circuit-level implementation details. For example, a multi-controlled NOT (X) gate can be created in various ways depending on the number of auxiliary qubits the implementation requires \cite{barenco95, Nielsen_Chuang_2010, PhysRevA.93.022311, 10.1145/3656436, Gidney2015}. Auxiliary qubits act as the quantum analog of classical temporary variables; they store intermediate quantum results instead of recomputing them, thereby increasing the circuit's execution speed.

\begin{figure}
    \centering
    \begin{tikzpicture}[>=stealth, node distance=1cm, every node/.style={circle, text centered}]
        \node (A) [draw=allocation_border, fill=allocation_fill] {A, 5};
        \node (B) [above right=of A, draw=neutral_border, fill=neutral_fill] {B};
        \node (C) [right=of A, draw=neutral_border, fill=neutral_fill] {C};
        \node (D) [below right=of A, draw=neutral_border, fill=neutral_fill] {D};
        \node (E) [right=of B, draw=neutral_border, fill=neutral_fill] {E};
        \node (F) [right=of C, draw=deallocation_border, fill=deallocation_fill] {F, 1};
        \node (G) [right=of D, draw=allocation_border, fill=allocation_fill] {G, 1};
        
        \draw[->] (A) -- (B);
        \draw[->] (A) -- (C);
        \draw[->] (A) -- (D);
        \draw[->] (B) -- (E);
        \draw[->] (A) -- (E);
        \draw[->] (C) -- (F);
        \draw[->] (D) -- (G);
        \draw[->] (C) -- (G);
    \end{tikzpicture}
    \caption{\justifying A control flow DAG corresponding to some quantum program. The nodes A and G are allocation nodes and are marked in \textcolor{allocation_border}{red}. The nodes B, C, D, and E are neutral and are marked in \textcolor{neutral_border}{blue}. The node F is a deallocation node and is marked in \textcolor{deallocation_border}{green}. The numbers correspond to the required or released qubit count.
    }
    \label{fig:reuse-dag}
\end{figure}
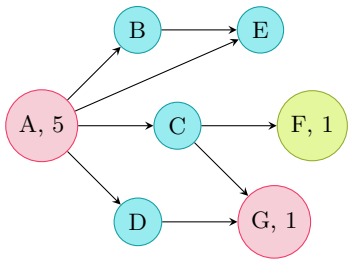

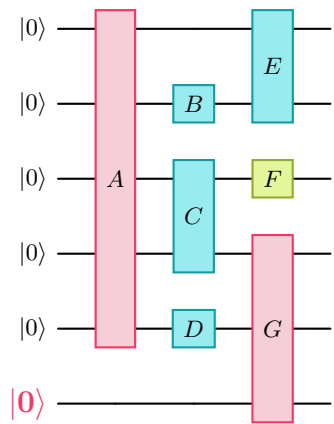
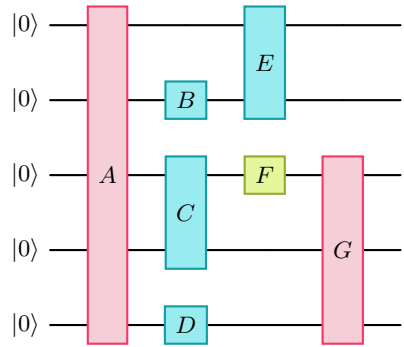
\begin{figure}
    \centering
    \begin{subfigure}{0.45\textwidth}
        \centering
        \begin{quantikz}
            \lstick{\ket{0}} & \gate[5, style={fill=allocation_fill, draw=allocation_border}]{A} & & \gate[2, style={fill=neutral_fill, draw=neutral_border}]{E} & \\
            \lstick{\ket{0}} & & \gate[1, style={fill=neutral_fill, draw=neutral_border}]{B} & & \\
            \lstick{\ket{0}} & & \gate[2, style={fill=neutral_fill, draw=neutral_border}]{C} & \gate[1, style={fill=deallocation_fill, draw=deallocation_border}]{F} & \\
            \lstick{\ket{0}} & & & \gate[3, style={fill=allocation_fill, draw=allocation_border}]{G} & \\
            \lstick{\ket{0}} & & \gate[1, style={fill=neutral_fill, draw=neutral_border}]{D} & & \\
            \lstick{\large \textcolor{allocation_border}{\ket{{\mathbf{0}}}}} & & & &
        \end{quantikz}
        \caption{}
        \label{subfig:no-reuse-applied}
    \end{subfigure}
    \begin{subfigure}{0.45\textwidth}
        \centering
        \begin{quantikz}
            \lstick{\ket{0}} & \gate[5, style={fill=allocation_fill, draw=allocation_border}]{A} & & \gate[2, style={fill=neutral_fill, draw=neutral_border}]{E} & & \\
            \lstick{\ket{0}} & & \gate[1, style={fill=neutral_fill, draw=neutral_border}]{B} & & & \\
            \lstick{\ket{0}} & & \gate[2, style={fill=neutral_fill, draw=neutral_border}]{C} & \gate[1, style={fill=deallocation_fill, draw=deallocation_border}]{F} & \gate[3, style={fill=allocation_fill, draw=allocation_border}]{G} & \\
            \lstick{\ket{0}} & & & & & \\
            \lstick{\ket{0}} & & \gate[1, style={fill=neutral_fill, draw=neutral_border}]{D} & & & \\
        \end{quantikz}
        \caption{}
        \label{subfig:reuse-applied}
    \end{subfigure}

    \caption{\justifying 
    Two different quantum circuit implementations for the control flow DAG in Figure \ref{fig:reuse-dag}.}
    \label{fig:reuse-circuits}
\end{figure}

Figures \ref{fig:reuse-dag}, \ref{fig:reuse-circuits}, \ref{fig:reuse-dag-auxiliaries}, and \ref{fig:reuse-circuits-auxiliaries}, present examples of control flow graphs and their corresponding quantum circuit implementations. Different implementations exhibit the trade-off between the total qubit count and the circuit's execution runtime. Reusing qubits favors the former, while avoiding it benefits the latter. Figure \ref{fig:reuse-dag} shows an example of a control flow graph with allocation and deallocation nodes, and Figure \ref{fig:reuse-circuits} shows its corresponding possible implementations. In \subref{subfig:no-reuse-applied}, the allocation node G is scheduled in parallel to the deallocation node F, requiring a new qubit from the pool and preventing possible reuse from the deallocation node F. In \subref{subfig:reuse-applied}, the deallocation node F is scheduled first, allowing its qubits to be reused by the second allocation node. This choice reduces qubit usage but may increase the circuit depth since the number of nodes connected serially increases to four, compared to three when qubits are not recycled. Figure \ref{fig:reuse-dag-auxiliaries} and Figure \ref{fig:reuse-circuits-auxiliaries}, respectively, provide a similar example, with an allocation node and a deallocation node replaced by nodes with auxiliary qubits. In \subref{subfig:no-reuse-applied-auxiliaries}, the node G is scheduled in parallel to the node F, requiring two new qubits from the pool to account for the auxiliary qubits. Reuse is prevented. In \subref{subfig:reuse-applied-f-g-auxiliaries}, the node F is scheduled first, allowing G to reuse its auxiliary qubit. The opposite scenario is also possible and has been omitted for brevity. As in Figure \ref{fig:reuse-circuits}, the different implementations exhibit the trade-off between qubit count and depth.

\begin{figure}
    \centering
    \begin{tikzpicture}[>=stealth, node distance=1cm, every node/.style={circle, text centered}]
        \node (A) [draw=allocation_border, fill=allocation_fill] {A, 5};
        \node (B) [above right=of A, draw=neutral_border, fill=neutral_fill] {B};
        \node (C) [right=of A, draw=neutral_border, fill=neutral_fill] {C};
        \node (D) [below right=of A, draw=neutral_border, fill=neutral_fill] {D};
        \node (E) [right=of B, draw=neutral_border, fill=neutral_fill] {E};
        \node (F) [right=of C, draw=neutral_border, fill=neutral_fill] {F};
        \node at (F) [below right=0.04cm] {aux, 1};
        \node (G) [right=of D, draw=neutral_border, fill=neutral_fill] {G};
        \node at (G) [below right=0.04cm] {aux, 1};

        \draw[->] (A) -- (B);
        \draw[->] (A) -- (C);
        \draw[->] (A) -- (D);
        \draw[->] (B) -- (E);
        \draw[->] (A) -- (E);
        \draw[->] (C) -- (F);
        \draw[->] (D) -- (G);
        \draw[->] (C) -- (G);
    \end{tikzpicture}
    \caption{\justifying A control flow DAG corresponding to a quantum program similar to \ref{fig:reuse-dag} up to differences in nodes F and G. The node A is an allocation node. The nodes B, C, D, and E are neutral. Nodes G and F are also neutral, but each has an additional auxiliary qubit.}
    \label{fig:reuse-dag-auxiliaries}
\end{figure}
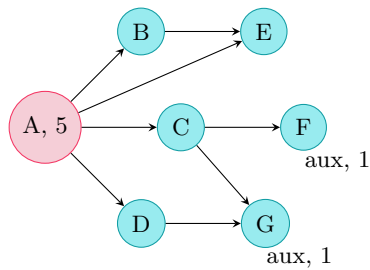

\begin{figure}
    \centering
    \begin{subfigure}{0.3\textwidth}
        \centering
        \begin{quantikz}
            \lstick{\ket{0}} & \gate[5, style={fill=allocation_fill, draw=allocation_border}]{A} & & \gate[2, style={fill=neutral_fill, draw=neutral_border}]{E} & [-0.85cm] & \\
            \lstick{\ket{0}} & & \gate[1, style={fill=neutral_fill, draw=neutral_border}]{B} & & & \\
            \lstick{\ket{0}} & & \gate[2, style={fill=neutral_fill, draw=neutral_border}]{C} & \gate[1, style={fill=neutral_fill, draw=neutral_border}]{F}\wire[d][3][style=dashed]{q} & & \\
            \lstick{\ket{0}} & & & & \gate[2, style={fill=neutral_fill, draw=neutral_border}]{G}\wire[d][3][style=dashed]{q} & \\
            \lstick{\ket{0}} & & \gate[1, style={fill=neutral_fill, draw=neutral_border}]{D} & & & \\
            \lstick{\large \textcolor{allocation_border}{\ket{{\mathbf{0}}}}} & & & \gate[1, style={fill=neutral_fill, draw=neutral_border}]{\text{aux}} & & \\
            \lstick{\large \textcolor{allocation_border}{\ket{{\mathbf{0}}}}} & & & & \gate[1, style={fill=neutral_fill, draw=neutral_border}]{\text{aux}} &
        \end{quantikz}
        \caption{}
        \label{subfig:no-reuse-applied-auxiliaries}
    \end{subfigure}
    \begin{subfigure}{0.3\textwidth}
        \centering
        \begin{quantikz}
            \lstick{\ket{0}} & \gate[5, style={fill=allocation_fill, draw=allocation_border}]{A} & & \gate[2, style={fill=neutral_fill, draw=neutral_border}]{E} & & \\
            \lstick{\ket{0}} & & \gate[1, style={fill=neutral_fill, draw=neutral_border}]{B} & & & \\
            \lstick{\ket{0}} & & \gate[2, style={fill=neutral_fill, draw=neutral_border}]{C} & \gate[1, style={fill=neutral_fill, draw=neutral_border}]{F}\wire[d][3][style=dashed]{q} & & \\
            \lstick{\ket{0}} & & & & \gate[2, style={fill=neutral_fill, draw=neutral_border}]{G}\wire[d][2][style=dashed]{q} & \\
            \lstick{\ket{0}} & & \gate[1, style={fill=neutral_fill, draw=neutral_border}]{D} & & & \\
            \lstick{\large \textcolor{allocation_border}{\ket{{\mathbf{0}}}}} & & & \gate[1, style={fill=neutral_fill, draw=neutral_border}]{\text{aux}} & \gate[1, style={fill=neutral_fill, draw=neutral_border}]{\text{aux}} & \\
        \end{quantikz}
        \caption{}
        \label{subfig:reuse-applied-f-g-auxiliaries}
    \end{subfigure}

    \caption{\justifying Three different quantum circuit implementations for the control flow DAG in Figure \ref{fig:reuse-dag-auxiliaries}.
    }
    \label{fig:reuse-circuits-auxiliaries}
\end{figure}
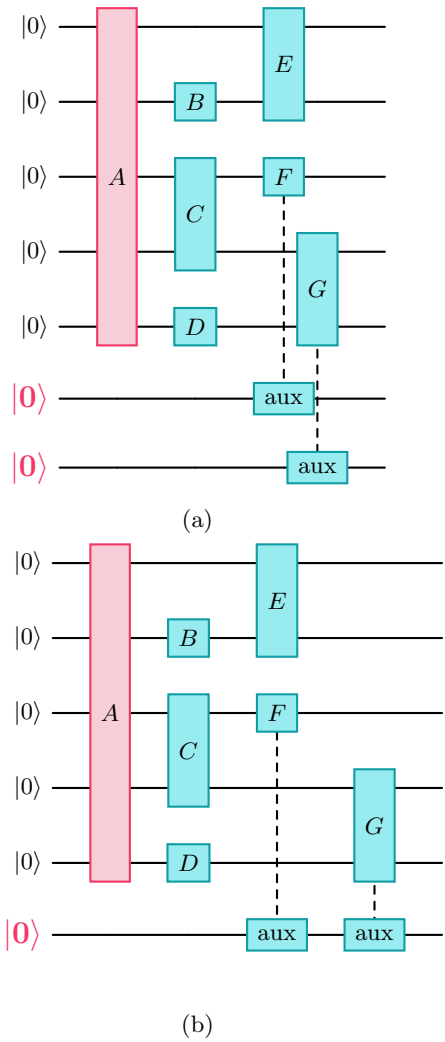

\section{Topological Sorting Analysis Pass}\label{sec:topological-sorting}

Our strategy for the analysis pass is to schedule as many qubit-releasing nodes as possible before applying qubit-requiring nodes. This leads to the greatest number of possible reuse options. Missing such an opportunity for reuse means that the scheduled qubit-requiring node will have to draw qubits from the device pool instead of recycling; this, in turn, increases the overall resource consumption of the program.

The scheduling process involves linearizing the graph, denoted by $G=\left<V,E\right>$, a process referred to as topological sorting \cite{kahn62}. This sorting preserves existing dependencies between nodes while fixing a specific order for independent nodes.

In Algorithm \ref{alg:smart-sort}, we generate a specific topological sort that prioritizes reuse options. It is based on the sorting of the qubit-releasing nodes. Thus, we generate a graph $G_{QR}=\left<V_{QR},E_{QR}\right>$ from $G$, where $V_{QR}$ is the set of qubit-releasing nodes and $\left(u,v\right)\in E_{QR}$ if $v$ is reachable from $u$ in $G$. This is the \textit{transitive closure} \cite{purdom_transitive_1970} of $G$ restricted to $V_{QR}$.

\begin{figure}
    \centering
    \begin{tikzpicture}[>=stealth, node distance=0.6cm, every node/.style={circle, text centered}]
        \node (A) [draw=allocation_border, fill=allocation_border] {};
        \node (B) [right=of A, draw=neutral_border, fill=neutral_border] {};
        \node (C) [above right=of B, draw=neutral_border, fill=neutral_border] {};
        \node (D) [below right=of B, draw=deallocation_border, fill=deallocation_border] {};
        \node (E) [above right=of C, draw=allocation_border, fill=allocation_border] {};
        \node (F) [above=0.3cm of E, draw=neutral_border, fill=neutral_border] {};
        \node (G) [above=0.3cm of F, draw=allocation_border, fill=allocation_border] {};
        \node (H) [below right=of D, draw=allocation_border, fill=allocation_border] {};
        \node (I) [above right=of D, draw=neutral_border, fill=neutral_border] {};
        \node (J) [right=of I, draw=allocation_border, fill=allocation_border] {};
        \node (K) [right=of H, draw=neutral_border, fill=neutral_border] {};
        \node (L) [right=of K, draw=deallocation_border, fill=deallocation_border] {};
        \node (M) [above right=of L, draw=neutral_border, fill=neutral_border] {};
        \node (N) [below right=of E, draw=deallocation_border, fill=deallocation_border] {};
        \node (O) [above right=of E, draw=neutral_border, fill=neutral_border] {};
        \node (P) [right=of N, draw=neutral_border, fill=neutral_border] {};
        \node (Q) [right=of P, draw=neutral_border, fill=neutral_border] {};
        \node (R) [right=of Q, draw=neutral_border, fill=neutral_border] {};
        \node (S) [right=of R, draw=neutral_border, fill=neutral_border] {};
        \node (T) [right=of S, draw=deallocation_border, fill=deallocation_border] {};
        \node (U) [right=of O, draw=allocation_border, fill=allocation_border] {};
        \node (V) [right=of U, draw=neutral_border, fill=neutral_border] {};

        \node (W) [below=1.5cm of A, draw=allocation_border, fill=allocation_border] {};
        \node (X) [right=of W, draw=neutral_border, fill=neutral_border] {};
        \node (Y) [right=of X, draw=deallocation_border, fill=deallocation_border] {};

        \node[draw, dashed, fit=(D), inner sep=0.1cm] {};
        \node[draw, dashed, fit=(N), inner sep=0.1cm] {};
        \node[draw, dashed, fit=(Y), inner sep=0.1cm] {};

        \node at (D) [below=0.15cm] {a};
        \node at (N) [below=0.1cm] {b};
        \node at (Y) [below=0.15cm] {c};
        \node at (A) [above left=0.1cm] {a, b};
        \node at (B) [above left=0.1cm] {a, b};
        \node at (C) [below=0.1cm] {b};
        \node at (E) [below=0.1cm] {b};
        \node at (X) [below=0.1cm] {c};
        \node at (W) [below=0.1cm] {c};
        \draw[->] (A) -- (B);
        \draw[->] (B) -- (C);
        \draw[->] (B) -- (D);
        \draw[->] (C) -- (E);
        \draw[->] (C) -- (F);
        \draw[->] (C) -- (G);
        \draw[->] (D) -- (H);
        \draw[->] (D) -- (I);
        \draw[->] (I) -- (J);
        \draw[->] (J) -- (M);
        \draw[->] (H) -- (K);
        \draw[->] (K) -- (L);
        \draw[->] (L) -- (M);
        \draw[->] (E) -- (N);
        \draw[->] (E) -- (O);
        \draw[->] (N) -- (P);
        \draw[->] (P) -- (Q);
        \draw[->] (Q) -- (R);
        \draw[->] (R) -- (S);
        \draw[->] (S) -- (T);
        \draw[->] (O) -- (U);
        \draw[->] (U) -- (V);
        \draw[->] (P) -- (V);
        \draw[->] (V) -- (T);
        \draw[->] (M) -- (R);
        \draw[->] (W) -- (X);
        \draw[->] (X) -- (Y);
    \end{tikzpicture}
    \caption{\justifying An example of a control flow DAG corresponding to an iteration in Algorithm \ref{alg:smart-sort}. 
    Allocation, neutral, and deallocation nodes are marked in \textcolor{allocation_border}{red}, \textcolor{neutral_border}{blue}, and \textcolor{deallocation_border}{green}, respectively.
    The nodes forming the qubit-releasing frontier are encircled by dashed lines. Each node has its corresponding ancestry. Each node and its ancestry are marked together with the letters a, b, and c.}
    \label{fig:topological-sorting-dag}
\end{figure}
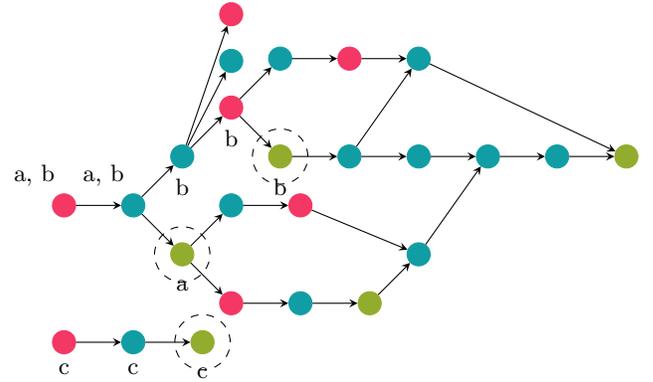

The algorithm iterates over $V_{QR}$ by dynamically maintaining the current frontier of $G_{QR}$, denoted by $F_{QR}$, and heuristically choosing the next node of the frontier, according to some cost function. For each node, we then obtain a topological sort of the subgraph containing the node and its predecessors, and append it to the full sort. By definition, the subgraph contains only one qubit-releasing node. Therefore, the explicit implementation of this sorting does not affect the reusability of the qubits. (An exception to this is nodes that have auxiliary qubits, an issue we address in Subsection \ref{subsec:aux-qubits}.) We then update $G_{QR}$ and $F_{QR}$ accordingly and continue the iteration. An example of such iteration is shown in Figure \ref{fig:topological-sorting-dag}.

By iterating over the qubit-releasing nodes first, we heuristically follow the algorithm's goal of scheduling qubit-releasing nodes early. We can select specific implementations of the topological sorting for each node's ancestry and the cost function to prioritize other aspects of the sorting. The total qubit requirement of the node's ancestry is the cost function we found to be simple and practically effective.

\begin{algorithm}[H]
\caption{Topological Sorting to Maximize Reuse Options}
\label{alg:smart-sort}
\begin{algorithmic}[1]
\Function{SMART\_SORT}{$G$}
    \State $G_{QR} \gets$ Transitive closure of $G$, restricted to $V_{QR}$
    \State $FULL\_SORT \gets []$
    \State $F_{QR} \gets$ Frontier of $G_{QR}$
    \While{$F_{QR} \neq \emptyset$}
        \State $v \gets $ node from $F_{QR}$ which minimizes $COST\left(v\right)$
        \State $ancestors \gets $ ancestry of $v$ in $G$
        \State Append $TOPOLOGICAL\_SORTING(ancestors)$ to $FULL\_SORT$
        \State Remove $v$ from $G_{QR}$
        \State Remove $ancestors$ from $G$
        \State Update $F_{QR}$
    \EndWhile
    \State Append $TOPOLOGICAL\_SORTING\left(G\right)$ to $FULL\_SORT$
\EndFunction
\end{algorithmic}
\end{algorithm}

\subsection{Complexity Analysis}

\noindent Claim: \textit{The complexity of the proposed algorithm with the cost function of the qubit requirement per ancestry is quadratic in the number of nodes $\left|V\right|$}. The following discussion outlines the reasoning for this claim.

We assume that the control flow graph is sparse, i.e., the average number of edges per node remains constant regardless of the size of the graph. Namely, $\mathcal{O}\left(|E|\right)=\mathcal{O}\left(|V|\right)$. We can make this assumption because the graph characterizes a scheduling of quantum operations. Therefore, the maximum number of edges per function is the number of I/O variables the function consumes, independent of the overall number of nodes in the graph.

During initialization, the transitive closure of the qubit-releasing nodes can be obtained in $\mathcal{O}\left(|V|\cdot|V_{QR}|\right)$ by applying a procedure similar to Purdom's algorithm \cite{purdom_transitive_1970}. This algorithm generates a topologically sorted array. Due to the graph's sparsity assumption, this step is of $\mathcal{O}\left(\left|V\right|\right)$. Then, it iterates over the array in reverse to form the accumulated transitive closure. We modify this step by keeping only the qubit-releasing nodes, reducing the complexity from $\mathcal{O}\left(\left|V\right|^2\right)$ to $\mathcal{O}\left(|V|\cdot|V_{QR}|\right)$.

Iterating over and updating $F_{QR}$ can be done using Kahn's algorithm \cite{kahn62} in $\mathcal{O}\left(|V_{QR}|\right)$. The complexity of finding the next node within $F_{QR}$ depends on the complexity of computing the cost function, which we denote by $f\left(n\right)$. Thus, we find the node that minimizes the cost function in $\mathcal{O}\left(|V_{QR}|\cdot f(n)\right)$. Finding the ancestors of a node, topologically sorting them, and removing them from $G$ can be done in $\mathcal{O}\left(|V|\right)$. However, since we do this just once for each node (and then remove it), we guarantee that each node and each edge require a constant number of operations over the whole iteration. Thus, the runtime is $\mathcal{O}\left(|V|+|V_{QR}|^2\cdot f\left(n\right)\right)$ for the iteration and  $\mathcal{O}\left(|V|\cdot|V_{QR}|+|V_{QR}|^2\cdot f\left(n\right)\right)$ for the entire algorithm.

Naively, the suggested cost function $f\left(n\right)$ introduces a linear factor of $\left|V\right|$ to the complexity because it iterates over each node's ancestry. However, the procedure can be extended to include two precomputed data structures that would avoid this runtime increase. The first corresponds to the ancestry of each qubit-releasing node and the cost function value, and the second involves the reverse mapping between the nodes and their qubit-releasing descendants. All steps are $\mathcal{O}\left(\left|V\right|\cdot\left|V_{QR}\right|\right)$ in runtime and memory since the ancestry of each qubit-releasing node can be computed in $\mathcal{O}\left(\left|V\right|\right)$. 
After each iteration, the ancestry of the chosen qubit-releasing node is removed from the graph. For each removed node, its qubit-releasing descendants are obtained from the precomputed array and the cost function is recalculated for these descendants. The cost function remains intact for the rest of the qubit-releasing nodes. Each node in the graph is only removed once, guaranteeing recalculations in $\mathcal{O}\left(\left|V\right|\cdot\left|V_{QR}\right|\right)$ over the whole algorithm. The result is an overall complexity of $\mathcal{O}\left(\left|V\right|\cdot\left|V_{QR}\right|\right)$ for the combined iteration and cost function calculation steps. 
To conclude, the complexity of the entire algorithm with our choice of cost function is $\mathcal{O}\left(\left|V\right|\cdot\left|V_{QR}\right|\right)$.

Practically, the number of qubit-releasing nodes grows linearly with the number of the overall nodes, resulting in a quadratic complexity for the algorithm $\mathcal{O}\left(\left|V\right|^2\right)$. From now on, we omit the subscript $QR$ from $V_{QR}$ and replace it by $V$.

\subsection{Handling Auxiliary Qubits} \label{subsec:aux-qubits}

We argue that the penalty for missing the reuse option for nodes with auxiliary qubits is not likely to be significant for circuits that are large enough. Even if an iteration of Algorithm \ref{alg:smart-sort} ordered an allocation node before a node with auxiliary qubits, the latter are reserved for reuse in subsequent iterations. Still, simple heuristics can introduce more reuse options and further optimize the circuit's resources. Algorithm \ref{alg:smart-sort} calls $TOPOLOGICAL\_SORTING$ in each iteration. This function is an adjustable black box. Here, it can prioritize nodes with auxiliary qubits over qubit-requiring nodes when possible, thereby allowing more reuse options.

\section{Reuse Analysis Pass}\label{sec:reuse-analysis-pass}

Once we obtain the topological sort, we can transform the intermediate representation of the control flow graph into a description that is closer to the circuit level and involves the connection between the qubit-releasing nodes and the qubit-requiring nodes. Qubit-requiring nodes that are not connected for reuse will draw qubits from the device pool.

The pseudocode in Algorithm \ref{alg:qubit-reuse-strategy} traverses the nodes according to the topological sort determined in the previous pass, and consumes and releases qubits on the fly. The algorithm includes a heuristic-based component to identify reuse options. For each node, the heuristic determines whether to reuse qubits or allocate qubits from the device pool. If it chooses the former, it decides which recycled qubits and required qubits are matched together. Different heuristics can prioritize optimizing different characteristics of the output circuit such as number of qubits and depth. Choosing the right heuristic depends on the user preferences, optimization level, and computational resources.

Each node must have a pre-determined number of auxiliary qubits. This means that the implementation of each function has already been selected. Alternatively, we can postpone the reuse assignment of the auxiliary qubits for a different pass because auxiliary qubits both consume and release qubits. This partial information, however, can lead to some performance drawbacks depending on the heuristics.

If the program constraints are tighter and require fine-tuning of both width and depth, we can replace this pass with more extensive search techniques such as backtracking.

\begin{algorithm}[H]
\caption{Apply Qubit Reuse}
\label{alg:qubit-reuse-strategy}
\begin{algorithmic}[1]
    \Function{QUBIT\_REUSE\_STRATEGY}{$topological\_sort$}
        \State $REUSE\_POOL \gets []$
        \ForAll{$v\in topological\_sort$}
            \If{$v$ is qubit-requiring or uses auxiliary qubits}
                \State
                \begin{flushleft}
                $RO \gets FIND\_REUSE\_OPTION\left(v, REUSE\_POOL\right)$
                \end{flushleft}
                \ForAll{$\left(v',j\right), \left(v,i\right) \in RO$}
                    \State Connect $\left(v',j\right)$ and $\left(v,i\right)$
                    \State Remove $\left(v',j\right)$ from $REUSE\_POOL$
                \EndFor
            \EndIf
            \State 
            \begin{flushleft}
            Append auxiliary and released qubits from $v$ to $REUSE\_POOL$
            \end{flushleft}
        \EndFor
    \EndFunction
\end{algorithmic}
\end{algorithm}

Algorithm \ref{alg:qubit-reuse-strategy} is a simple iteration over the topological sort and scales linearly in the number of nodes. The runtime also scales linearly with the size of the reuse pool $Q$. Thus, the complexity of the pass is $\mathcal{O}\left(\left(Q + f\left(Q\right)\right)\cdot|V|\right)$, where $f$ is the complexity of $FIND\_REUSE\_OPTION$. $Q$ is bounded by the number of possible qubits that can be released. This number scales linearly with the number of qubit-releasing nodes, including those with auxiliary qubits, and $\bar{Q}$, which represents the qubit count per node. In many practical cases, the size of the reuse pool is significantly smaller and less likely to cause a bottleneck.

\subsection{Reuse Strategies}

We now review various heuristics for deciding how many qubits to reuse and which ones to select. Each heuristic is influenced by the algorithm's execution preferences and balances the trade-off between circuit width and depth, the latter corresponding to the execution runtime. The method straightforwardly supports gate-specific depth metrics such as the T-depth or CX-depth that characterize circuit quality for Fault Tolerant or NISQ hardware, respectively.

\subsubsection{Greedy Reuse to Optimize Width}

The strategy for optimization using greedy reuse seeks to minimize the overall number of qubits consumed by the circuit. Every reuse opportunity is selected immediately. The order of selection does not affect the reuse effectiveness as long as reuse is being done, because any qubit that was not chosen for reuse in the current step could be reused later. We get $f\left(Q\right)=\mathcal{O}\left(Q\right)$ and the overall complexity of the pass is $\mathcal{O}\left(Q\cdot\left|V\right|\right)$. For higher optimization levels, a more advanced heuristic would consider the optimal qubit to choose for reuse. For example, a relatively naive heuristic to minimize depth would choose the qubit with the smallest number of operations applied to it. Clearly, this may affect the complexity of the pass.

\subsubsection{Dependency-Preserving Reuse}\label{subsec:dependency-preserving}

Here, we allow function B to reuse the qubits of function A, if B is a descendant of A. On the other hand, functions that run in parallel will not share auxiliary qubits between them. In most reasonable cases, this will ensure that the circuit’s depth does not grow due to the reuse operations, though this may come at the potential cost of missing reuse opportunities and increasing the circuit's width. By pre-computing the transitive closure of the graph, we can check the relation between a given pair of functions in constant time; the overall complexity of the pass is $\mathcal{O}\left(Q\cdot\left|V\right|\right)$.

\subsubsection{Depth-Preserving Reuse}

The heuristic used to prevent an increase in circuit depth, shown in Algorithm \ref{alg:reuse-option-no-depth-increase}, takes a more permissive approach than the previous one. It allows parallel functions to reuse qubits if the depth of the circuit does not increase as a result. This complex optimization results in a longer compilation runtime. The heuristic tracks the increase in the depth of the circuit as we "append" more nodes. We allow function B to reuse qubits from function A if adding an edge between the corresponding nodes does not increase the depth.

The accumulated depth of the circuit at a specific qubit refers to the total time steps required to execute all operations on this qubit (including their dependencies). We compute it conservatively, without considering potential gate cancellations between adjacent operations. This keeps the calculation tractable while keeping the result within the same order of magnitude, as these cancellations are expected to have a weaker impact. 

Computing the accumulated depth requires processing each node's gate-level implementation.\footnote{Some nodes may be compound, representing multiple operations. In this case, we would be required to either pre-compile them to get the gate-level implementations, or use estimations rather than exact values.} This is computationally expensive, so we perform some of these computations as a preprocessing stage. Each node is represented as a DAG, similar to the control flow graph described earlier, but at a lower level, considering basic gates and qubits. This representation is available in the Qiskit\cite{qiskit} and TKET\cite{tket} quantum Software Developer Kits (SDKs). In this DAG, each operation is represented by a vertex and the edges represent the data flow between operations. Each qubit is represented by an input vertex, connected to the first operation applied to the qubit, and an output vertex, connected to the last operation applied to the qubit. We use this representation to compute $DEP\left(v,i,j\right)$, the shortest path between the input vertex of the qubit $i$ to the output vertex of qubit $j$ in this DAG. At this stage, we can also take into account low-level optimizations and the target architecture's native gate set to improve our depth computation.\footnote{In partially connected hardware, accurate depth computation requires knowledge of the physical mapping to account for SWAP operations. Taking this into account requires a more complex algorithm. We discuss connectivity further in Section \ref{sec:future-work}.}

While iterating over the topological sort, we keep track of the accumulated depth of the circuit at each qubit. This is done by computing $PRE\_DEP\left(v,i\right)$ ($POST\_DEP\left(v,i\right)$), which represents the depth of the circuit composed of all nodes up to and excluding (including) $v$ at qubit $i$. After we compute this value, we iterate over all qubits in the pool and check which ones will not increase the accumulated depth. Of these, we heuristically choose the qubit with the maximal $POST\_DEP$. The rationale behind this ranking is that this qubit does not increase the circuit's depth at this time. That said, it may increase the depth in the next iterations. Therefore, eliminating it now prevents a higher depth penalty later.

\begin{algorithm}[H]
\caption{Find Reuse Option w/o Increasing Depth}
\label{alg:reuse-option-no-depth-increase}
\begin{algorithmic}[1]
    \Function{FIND\_REUSE\_OPTION}{$v,REUSE\_POOL$}
        \ForAll{$i\in$ qubits of $v$}
            \If{$i$ is taken from the auxiliary pool}
                \State $PRE\_DEP\left(i\right) = 0$
            \Else
                \State $v' \gets$ the previous node applied to $i$
                \State $PRE\_DEP\left(i\right)=POST\_DEP\left(v',i\right)$
                \State \Comment{$POST\_DEP\left(v',i\right)$ is guaranteed to be known since the iteration is done in topological order.}
            \EndIf
        \EndFor

        \ForAll{$i\in$ qubits of $v$}
            \State
            \begin{flushleft} 
            $POST\_DEP\left(v,i\right) =
            \max_j {\left(PRE\_DEP\left(j\right) + DEP\left(v,j,i\right)\right)}$
            \end{flushleft}
        \EndFor
        \vspace{1mm}
    
        \State $REUSE\_OPTION \gets []$
        \ForAll{$i\in$ requiring qubits of $v$}
            \State 
            \begin{flushleft}
            $candidates\gets$ all $\left(v',j\right)\in REUSE\_POOL$ such that \\
            $POST\_DEP\left(v',j\right)+DEP\left(v,i,k\right) \leq
            POST\_DEP\left(v,k\right)$ \\
            \textbf{for all} $k \in$ qubits of $v$
            \end{flushleft}
            \Statex
            \If{$candidates\neq\emptyset$}
                \State 
                \begin{flushleft}
                Choose $\left(v',j\right)\in candidates$ that maximize $POST\_DEP\left(v',j\right)$
                \end{flushleft}
                \State Append $\left(v',j\right), \left(v,i\right)$ to $REUSE\_OPTION$
            \EndIf
        \EndFor
        \State \Return{$REUSE\_OPTION$}
    \EndFunction
\end{algorithmic}
\end{algorithm}

As described above, computing $DEP\left(v,i,j\right)$ requires processing the gate-level implementation of each node and performing optimization steps, possibly taking into account the target architecture. These computations are performed for each node separately, rather than the whole circuit. Still, this can quickly become a bottleneck, depending on the optimization and accuracy level required. We omit the computation of this low-level complexity analysis, as it is beyond the scope of this paper and there is extensive research on this topic \cite{iten22, nam18, tket}.

For a given qubit, computing $POST\_DEP$ requires iterating over all qubits, resulting in quadratic complexity in the number of qubits. Overall, the complexity of this calculation is $\mathcal{O}\left(\bar{Q}^{2}\right)$ and for the entire pass that iterates over all graph nodes it is given by $\mathcal{O}\left(\bar{Q}^{2}\cdot\left|V\right|\right)$. 

We determine the reuse option for the qubit-requiring nodes, including auxiliary nodes. This requires iterating over the qubit-releasing pool to determine valid candidates by comparing how much they would influence the current node's depth. As mentioned previously, the pool has a worst-case size of $Q$, while the comparison step provides a factor of $\bar{Q}$. 
Thus, the overall complexity of the pass is $\mathcal{O}\left(Q\cdot\bar{Q}^{2}\cdot\left|V\right|\right)$, and the worst case scenario is $\mathcal{O}\left(\bar{Q}^{3}\cdot\left|V\right|^2\right)$. 

\section{Scalability} \label{sec:scalability}

The topological sorting algorithm grows quadratically with the number of nodes, which could lead to a compilation bottleneck in large circuits.

One approach to mitigating the quadratic growth could automatically or manually identify repeated code parts that use auxiliary qubits during compilation, pre-compile them internally according to the above algorithm, and connect them.

\begin{algorithm}[H]
\begin{algorithmic}[1]
    \Function{FOO}{}
        \State Initialize: $aux \gets |0\rangle$
        \State $\cdots$
        \State Uncompute: $aux \gets |0\rangle$
    \EndFunction

    \Loop
        \State FOO()
    \EndLoop
\end{algorithmic}
\end{algorithm}

\begin{figure}
    \centering
    \begin{subfigure}{0.4\textwidth}
        \centering
        \begin{tikzpicture}[>=stealth, node distance=0.3cm, every node/.style={circle, text centered}]
            \node (foo_0) [rectangle, minimum height=1cm, minimum width=0.5cm, draw=neutral_border, fill=neutral_fill] {};
            \node (all_0) [below left=of foo_0, draw=allocation_border, fill=allocation_border] {};
            \node (deall_0) [below right=of foo_0, draw=deallocation_border, fill=deallocation_border] {};
            \node (foo_1) [right=1.5cm of foo_0, rectangle, minimum height=1cm, minimum width=0.5cm, draw=neutral_border, fill=neutral_fill] {};
            \node (all_1) [below left=of foo_1, draw=allocation_border, fill=allocation_border] {};
            \node (deall_1) [below right=of foo_1, draw=deallocation_border, fill=deallocation_border] {};
            \node (foo_2) [right=1.5cm of foo_1, rectangle, minimum height=1cm, minimum width=0.5cm, draw=neutral_border, fill=neutral_fill] {};
            \node (all_2) [below left=of foo_2, draw=allocation_border, fill=allocation_border] {};
            \node (deall_2) [below right=of foo_2, draw=deallocation_border, fill=deallocation_border] {};
            \node (end) [right=1cm of foo_2, draw=none, fill=none] {};
            \node (dashing_0) [draw, dashed, rectangle, fit=(foo_0) (all_0) (deall_0)] {};
            \node[above=0.01cm of dashing_0] {FOO};
            
            \draw[->] (all_0) -- (foo_0);
            \draw[->] (foo_0) -- (deall_0);
            \draw[->] (all_1) -- (foo_1);
            \draw[->] (foo_1) -- (deall_1);
            \draw[->] (all_2) -- (foo_2);
            \draw[->] (foo_2) -- (deall_2);
            \draw[->] (foo_0) -- (foo_1);
            \draw[->] (foo_1) -- (foo_2);
            \draw[->, dashed] (foo_2) -- (end);
        \end{tikzpicture}
        \caption{}
        \label{subfig:repeated-foo-dag}
    \end{subfigure}
    \begin{subfigure}{0.4\textwidth}
        \centering
        \begin{tikzpicture}[>=stealth, node distance=0.3cm, every node/.style={circle, text centered}]
            \node (foo_0) [rectangle, minimum height=1cm, minimum width=0.5cm, draw=neutral_border, fill=neutral_fill] {};
            \node (all_0) [below left=of foo_0, draw=allocation_border, fill=allocation_border] {};
            \node (deall_0) [below right=of foo_0, draw=deallocation_border, fill=deallocation_border] {};
            \node (foo_1) [right=1.5cm of foo_0, rectangle, minimum height=1cm, minimum width=0.5cm, draw=neutral_border, fill=neutral_fill] {};
            \node (all_1) [below left=of foo_1, draw=allocation_border, fill=allocation_border] {};
            \node (deall_1) [below right=of foo_1, draw=deallocation_border, fill=deallocation_border] {};
            \node (foo_2) [right=1.5cm of foo_1, rectangle, minimum height=1cm, minimum width=0.5cm, draw=neutral_border, fill=neutral_fill] {};
            \node (all_2) [below left=of foo_2, draw=allocation_border, fill=allocation_border] {};
            \node (deall_2) [below right=of foo_2, draw=deallocation_border, fill=deallocation_border] {};
            \node (end) [right=1cm of foo_2, draw=none, fill=none] {};
            \node (end_deall) [right=1cm of deall_2, draw=none, fill=none] {};
            
            \draw[->] (all_0) -- (foo_0);
            \draw[->] (foo_0) -- (deall_0);
            \draw[->] (all_1) -- (foo_1);
            \draw[->] (foo_1) -- (deall_1);
            \draw[->] (all_2) -- (foo_2);
            \draw[->] (foo_2) -- (deall_2);
            \draw[->] (foo_0) -- (foo_1);
            \draw[->] (foo_1) -- (foo_2);
            \draw[->, dashed] (foo_2) -- (end);
            \draw[->] (deall_0) -- (all_1);
            \draw[->] (deall_1) -- (all_2);
            \draw[->, dashed] (deall_2) -- (end_deall);
        \end{tikzpicture}
        \caption{}
        \label{subfig:repeated-foo-dag-blocks}
    \end{subfigure}
    \caption{\justifying Control flow DAGs corresponding to the repeated application of $FOO$ in Section \ref{sec:scalability}. Figure \subref{subfig:repeated-foo-dag} includes the graph without modifications. Each \textcolor{neutral_border}{blue} node corresponds to $FOO$'s body between the allocation and deallocation statements, and the dashing surrounds the entire function.  In Figure \subref{subfig:repeated-foo-dag-blocks} edges are added between all adjacent deallocation and allocation nodes.}
    \label{fig:repeated-foo}
\end{figure}
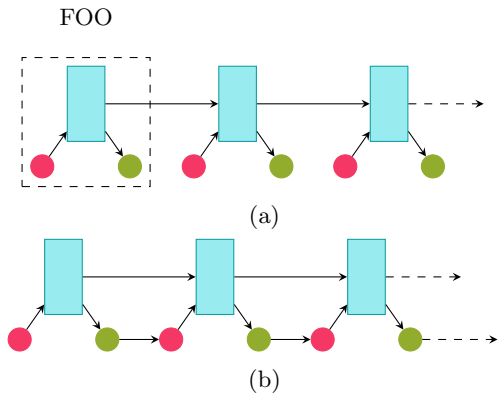

In this case, $FOO$ consists of an allocation node and a deallocation node. Inlining the code results in a series of allocation and deallocation nodes, as shown in Figure \ref{subfig:repeated-foo-dag}.

The formed graph has a large number of topological sorts. The first node in the sort can be selected out of $N$ nodes, but only the leftmost choice will not result in a qubit leak. After selecting the first node, we pick the nearest deallocation node and connect it to the second-leftmost node. This process continues sequentially. Intuitively, there is only a single sort that is sensible and does not cause the pool to bleed qubits: the one where the allocation and deallocation nodes are connected in series according to their associated iteration of $FOO$, shown in Figure \ref{subfig:repeated-foo-dag-blocks}. However, arriving at this conclusion requires substantial effort during the reuse pass. In this step, all qubit-releasing nodes are at the qubit-releasing frontier, and the correct order is determined based entirely on the choice of cost function.

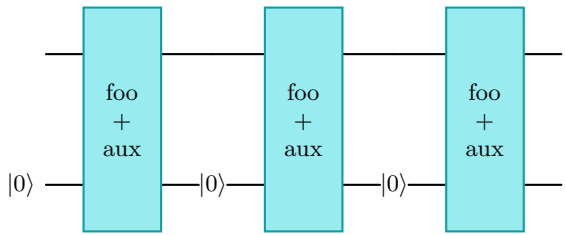
\begin{figure}
    \centering
    \begin{quantikz}
        & \gate[2, style={fill=neutral_fill, draw=neutral_border}]{\parbox{0.75cm}{foo \\ + \\ aux}} && \gate[2, style={fill=neutral_fill, draw=neutral_border}]{\parbox{0.75cm}{foo \\ + \\ aux}} && \gate[2, style={fill=neutral_fill, draw=neutral_border}]{\parbox{0.75cm}{foo \\ + \\ aux}} & \\
        \lstick{\ket{0}} & \push{\ket{0}} & \push{\ket{0}} &  & \push{\ket{0}} &  & \\
    \end{quantikz}
    \caption{\justifying A circuit implementation corresponding to \ref{fig:repeated-foo}, where the auxiliary qubits, marked with the state $\left|0\right>$, are reused sequentially, as dictated by \ref{subfig:repeated-foo-dag-blocks}.
    }
    \label{fig:repeated-foo-circuit}
\end{figure}

We could easily deduce the correct topological sort by analyzing the hierarchical information related to the nodes and realizing that each allocation-deallocation pair belongs to one iteration of $FOO$. As shown in Figure \ref{fig:repeated-foo-circuit}, The composite control flow graph, which is at a granularity level where the different iterations of $FOO$ are represented as nodes, has only one possible sort. Thus, we can reduce the runtime of the sorting to be quadratic in the size of the body of $FOO$ rather than the whole graph. The compiler is only required to apply the reuse pass to connect the allocation and deallocation pairs. Each iteration of $FOO$ is neutral in terms of qubit allocation, and these pairs act as auxiliary qubits.

More generally, algorithms may be partitioned into different sequences of allocating nodes, neutral (functional) nodes, and deallocating nodes, with the restriction imposed in Section \ref{sec:control-flow-graph} that a node cannot simultaneously draw and release non-overlapping qubits. Sorting each sequence individually, composing each subgraph into a node, and solving the composed problem can offer a significant runtime improvement, though with a possible trade-off in optimization. We now address the runtime benefits.

Consider a graph with $N$ nodes, partitioned into $P$ parts of equal size. Since the complexity of the topological sorting pass is quadratic, the sorting of each partition of the graph runs in $\mathcal{O}\left(\left(\frac{N}{P}\right)^2\right)$, and it runs $P$ times. The solutions are then composed into one graph with $P$ nodes, whose sorting runs in $\mathcal{O}\left(P^2\right)$. The overall runtime of the algorithm is then $\mathcal{O}\left(\frac{N^2}{P}+P^2\right)$. This expression exhibits a minimum at $P\sim N^{\frac{2}{3}}$, and the resulting complexity is reduced to $\mathcal{O}\left(N^{\frac{4}{3}}\right)$ at this optimal $P$. 

\begin{figure}
    \centering
    \begin{subfigure}{0.23\textwidth}
        \begin{tikzpicture}[>=stealth, node distance=1cm, every node/.style={circle, text centered}]
            \node (A) [draw=neutral_border, fill=neutral_fill] {A};
            \node (B) [above right=of A, draw=allocation_border, fill=allocation_fill] {B};
            \node (C) [below right=of A, draw=deallocation_border, fill=deallocation_fill] {C};
            \node[draw, ellipse, dashed, minimum width=3cm, minimum height=1.5cm, rotate=45] at ($(A)!0.5!(B)$) {};

            \draw[->] (A) -- (B);
            \draw[->] (A) -- (C);
        \end{tikzpicture}
        \caption{}
        \label{subfig:composite-dag-formation}
    \end{subfigure}
    \hfill
    \begin{subfigure}{0.23\textwidth}
        \begin{tikzpicture}[>=stealth, node distance=1cm, every node/.style={circle, text centered}]
            \node (AB) [ellipse, minimum width=2cm, minimum height=1cm, above right=of A, draw=allocation_border, fill=allocation_fill] {A + B};
            \node (C) [below right=of AB, draw=deallocation_border, fill=deallocation_fill] {C};
            
            \draw[->] (AB) -- (C);
        \end{tikzpicture}
        \caption{}
        \label{subfig:composite-dag}
    \end{subfigure}
    \caption{\justifying An example where composing arbitrary nodes results in loss of reuse options.
    }
    \label{fig:composite-topological-sorting}
\end{figure}
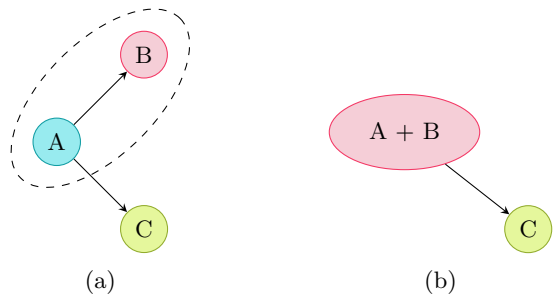

However, implementing an arbitrary partition into subgraphs could be detrimental to the reuse options. Figure \ref{fig:composite-topological-sorting} shows a simple example where considering the entire graph introduces more reuse options, contrary to the selected partitioning of the graph. The combination of nodes marked with $A$ and $B$, a neutral and an allocation one, respectively, shown in Figure \subref{subfig:composite-dag-formation}, results in the composite DAG in \subref{subfig:composite-dag}, where the combined node $A+B$ cannot reuse the qubits of the deallocation node $C$. This is the case even though such reuse is possible because node $C$ does not depend on node $B$. To address this difficulty, it is possible to introduce pattern detection schemes on the graph to optimize the partitions. Nevertheless, such detection may require significant additional computation. Implementing this algorithm in a scalable manner requires fine-tuning this balance based on the size of the input, user preferences, and compute resources. 

We can also consider a different method, following the hint from the example at the beginning of this section. A quantum program is formulated as a series of function calls, where some functions use local temporary variables that allocate and deallocate memory. We utilize this code structure to suggest a bottom-up approach, where the composition of the nodes is determined according to their functional association. The method is also allowed to inline certain function calls to get closer to the desired partition size of $P\sim N^{\frac{2}{3}}$. However, we practically found that the mere composition of a few function code blocks can dramatically improve the compilation runtime even without aiming for a specific partition size. Our optimization highlights the advantages of writing a quantum program using a high-level description. It requires maintaining high-level information after lowering the code to intermediate and low-level descriptions. Note that the partition should be determined after certain intermediate-level code optimizations that may affect temporary variables such as \cite{meuli19, paradis21, 10.1145/3656397, 10.1109/ISCA45697.2020.00054}. 

\section{Related Work}\label{sec:related-work}

Several existing SDK solutions provide limited support for qubit reuse, such as \cite{qiskit, cuda}. These solutions are primarily focused on multi-controlled NOT quantum operations that utilize auxiliary qubits and are applied sequentially. In contrast, our approach generalizes to any quantum unitary operation with auxiliary qubits, supports quantum operations that allocate or release qubits, and identifies reuse opportunities in modular quantum programs with independently scheduled functions.

Other studies \cite{10.1109/ISCA45697.2020.00054, seidel24} tackle qubit reuse in non-flat quantum programs. The method of Ding \textit{et al.} \cite{10.1109/ISCA45697.2020.00054} dynamically reclaims qubits after uncomputation, but depends on the order of function calls. In contrast, our approach is independent of function call order, enabling a broader exploration of reuse options. Seidel \cite{seidel24} employs topological sorting and a \textit{permeability graph} to expand the range of permissible sorts. However, our method introduces a broader spectrum of qubit reuse strategies and optimization levels, significantly enhancing scalability and making it better suited for managing qubits in larger, more complex quantum circuits.

\section{Possible Extensions and Future Work} \label{sec:future-work}

Qubit reuse strategies introduce different solutions to the trade-off between the number of qubits in a circuit and its depth. Handling this trade-off becomes more complicated when the target architecture only has partial connectivity. Namely, some reuse operations could introduce a significant routing overhead compared to others. This issue becomes even more pronounced when the target architecture has multiple quantum processing units, where different parts of the program might be scheduled in different units and the communication between them is expensive. The mapping of logical qubits to physical qubits and the routing are done later in the compilation pipeline, making it difficult to assess the effect of different reuse operations.

Algorithms that optimize qubit reuse can leverage the notion of "distance" between logical qubits. Qubits are considered close together if they need to be mapped to physically adjacent qubits due to the application of two-qubit gates between them or interactions involving their "neighboring" qubits. If there are too many close qubits, more SWAP operations are required. Thus, the algorithm could prioritize reuse options that do not decrease the distance between qubits. More permissive algorithms could allow qubits to be closer, but would need to avoid creating big "clusters" of qubits that are close together, e.g., limiting the size of a cluster to the size of each processing unit.

\section{Summary}
\label{sec:summary}

We presented a method for automatically reusing qubits that end up in the zero state after uncomputation, thereby reducing the qubit count required by the quantum program. The technique employs heuristic topological sorting of the control flow graph and leverages high-level information preservation to achieve scalability for large programs.

\bibliography{main}

\begin{thebibliography}{31}%
\makeatletter
\providecommand \@ifxundefined [1]{%
 \@ifx{#1\undefined}
}%
\providecommand \@ifnum [1]{%
 \ifnum #1\expandafter \@firstoftwo
 \else \expandafter \@secondoftwo
 \fi
}%
\providecommand \@ifx [1]{%
 \ifx #1\expandafter \@firstoftwo
 \else \expandafter \@secondoftwo
 \fi
}%
\providecommand \natexlab [1]{#1}%
\providecommand \enquote  [1]{``#1''}%
\providecommand \bibnamefont  [1]{#1}%
\providecommand \bibfnamefont [1]{#1}%
\providecommand \citenamefont [1]{#1}%
\providecommand \href@noop [0]{\@secondoftwo}%
\providecommand \href [0]{\begingroup \@sanitize@url \@href}%
\providecommand \@href[1]{\@@startlink{#1}\@@href}%
\providecommand \@@href[1]{\endgroup#1\@@endlink}%
\providecommand \@sanitize@url [0]{\catcode `\\12\catcode `\$12\catcode `\&12\catcode `\#12\catcode `\^12\catcode `\_12\catcode `\%12\relax}%
\providecommand \@@startlink[1]{}%
\providecommand \@@endlink[0]{}%
\providecommand \url  [0]{\begingroup\@sanitize@url \@url }%
\providecommand \@url [1]{\endgroup\@href {#1}{\urlprefix }}%
\providecommand \urlprefix  [0]{URL }%
\providecommand \Eprint [0]{\href }%
\providecommand \doibase [0]{https://doi.org/}%
\providecommand \selectlanguage [0]{\@gobble}%
\providecommand \bibinfo  [0]{\@secondoftwo}%
\providecommand \bibfield  [0]{\@secondoftwo}%
\providecommand \translation [1]{[#1]}%
\providecommand \BibitemOpen [0]{}%
\providecommand \bibitemStop [0]{}%
\providecommand \bibitemNoStop [0]{.\EOS\space}%
\providecommand \EOS [0]{\spacefactor3000\relax}%
\providecommand \BibitemShut  [1]{\csname bibitem#1\endcsname}%
\let\auto@bib@innerbib\@empty
\bibitem [{\citenamefont {Bennett}(1973)}]{5391327}%
  \BibitemOpen
  \bibfield  {author} {\bibinfo {author} {\bibfnamefont {C.~H.}\ \bibnamefont {Bennett}},\ }\bibfield  {title} {\bibinfo {title} {Logical reversibility of computation},\ }\href {https://doi.org/10.1147/rd.176.0525} {\bibfield  {journal} {\bibinfo  {journal} {IBM Journal of Research and Development}\ }\textbf {\bibinfo {volume} {17}},\ \bibinfo {pages} {525} (\bibinfo {year} {1973})}\BibitemShut {NoStop}%
\bibitem [{\citenamefont {Nielsen}\ and\ \citenamefont {Chuang}(2010)}]{Nielsen_Chuang_2010}%
  \BibitemOpen
  \bibfield  {author} {\bibinfo {author} {\bibfnamefont {M.~A.}\ \bibnamefont {Nielsen}}\ and\ \bibinfo {author} {\bibfnamefont {I.~L.}\ \bibnamefont {Chuang}},\ }\href@noop {} {\emph {\bibinfo {title} {Quantum Computation and Quantum Information: 10th Anniversary Edition}}}\ (\bibinfo  {publisher} {Cambridge University Press},\ \bibinfo {year} {2010})\BibitemShut {NoStop}%
\bibitem [{\citenamefont {Rand}\ \emph {et~al.}(2019)\citenamefont {Rand}, \citenamefont {Paykin}, \citenamefont {Lee},\ and\ \citenamefont {Zdancewic}}]{Rand_2019}%
  \BibitemOpen
  \bibfield  {author} {\bibinfo {author} {\bibfnamefont {R.}~\bibnamefont {Rand}}, \bibinfo {author} {\bibfnamefont {J.}~\bibnamefont {Paykin}}, \bibinfo {author} {\bibfnamefont {D.-H.}\ \bibnamefont {Lee}},\ and\ \bibinfo {author} {\bibfnamefont {S.}~\bibnamefont {Zdancewic}},\ }\bibfield  {title} {\bibinfo {title} {Reqwire: Reasoning about reversible quantum circuits},\ }\href {https://doi.org/10.4204/eptcs.287.17} {\bibfield  {journal} {\bibinfo  {journal} {Electronic Proceedings in Theoretical Computer Science}\ }\textbf {\bibinfo {volume} {287}},\ \bibinfo {pages} {299–312} (\bibinfo {year} {2019})}\BibitemShut {NoStop}%
\bibitem [{\citenamefont {Bennett}(1989)}]{doi:10.1137/0218053}%
  \BibitemOpen
  \bibfield  {author} {\bibinfo {author} {\bibfnamefont {C.~H.}\ \bibnamefont {Bennett}},\ }\bibfield  {title} {\bibinfo {title} {Time/space trade-offs for reversible computation},\ }\href {https://doi.org/10.1137/0218053} {\bibfield  {journal} {\bibinfo  {journal} {SIAM Journal on Computing}\ }\textbf {\bibinfo {volume} {18}},\ \bibinfo {pages} {766} (\bibinfo {year} {1989})},\ \Eprint {https://arxiv.org/abs/https://doi.org/10.1137/0218053} {https://doi.org/10.1137/0218053} \BibitemShut {NoStop}%
\bibitem [{\citenamefont {Meuli}\ \emph {et~al.}(2019)\citenamefont {Meuli}, \citenamefont {Soeken}, \citenamefont {Roetteler}, \citenamefont {Bjorner},\ and\ \citenamefont {Micheli}}]{meuli19}%
  \BibitemOpen
  \bibfield  {author} {\bibinfo {author} {\bibfnamefont {G.}~\bibnamefont {Meuli}}, \bibinfo {author} {\bibfnamefont {M.}~\bibnamefont {Soeken}}, \bibinfo {author} {\bibfnamefont {M.}~\bibnamefont {Roetteler}}, \bibinfo {author} {\bibfnamefont {N.}~\bibnamefont {Bjorner}},\ and\ \bibinfo {author} {\bibfnamefont {G.~D.}\ \bibnamefont {Micheli}},\ }\href {https://arxiv.org/abs/1904.02121} {\bibinfo {title} {Reversible pebbling game for quantum memory management}} (\bibinfo {year} {2019}),\ \Eprint {https://arxiv.org/abs/1904.02121} {arXiv:1904.02121 [quant-ph]} \BibitemShut {NoStop}%
\bibitem [{\citenamefont {Paradis}\ \emph {et~al.}(2021)\citenamefont {Paradis}, \citenamefont {Bichsel}, \citenamefont {Steffen},\ and\ \citenamefont {Vechev}}]{paradis21}%
  \BibitemOpen
  \bibfield  {author} {\bibinfo {author} {\bibfnamefont {A.}~\bibnamefont {Paradis}}, \bibinfo {author} {\bibfnamefont {B.}~\bibnamefont {Bichsel}}, \bibinfo {author} {\bibfnamefont {S.}~\bibnamefont {Steffen}},\ and\ \bibinfo {author} {\bibfnamefont {M.}~\bibnamefont {Vechev}},\ }\bibfield  {title} {\bibinfo {title} {Unqomp: synthesizing uncomputation in quantum circuits},\ }in\ \href {https://doi.org/10.1145/3453483.3454040} {\emph {\bibinfo {booktitle} {Proceedings of the 42nd ACM SIGPLAN International Conference on Programming Language Design and Implementation}}},\ \bibinfo {series and number} {PLDI 2021}\ (\bibinfo  {publisher} {Association for Computing Machinery},\ \bibinfo {address} {New York, NY, USA},\ \bibinfo {year} {2021})\ p.\ \bibinfo {pages} {222–236}\BibitemShut {NoStop}%
\bibitem [{\citenamefont {Paradis}\ \emph {et~al.}(2024)\citenamefont {Paradis}, \citenamefont {Bichsel},\ and\ \citenamefont {Vechev}}]{paradis24}%
  \BibitemOpen
  \bibfield  {author} {\bibinfo {author} {\bibfnamefont {A.}~\bibnamefont {Paradis}}, \bibinfo {author} {\bibfnamefont {B.}~\bibnamefont {Bichsel}},\ and\ \bibinfo {author} {\bibfnamefont {M.}~\bibnamefont {Vechev}},\ }\bibfield  {title} {\bibinfo {title} {Reqomp: Space-constrained uncomputation for quantum circuits},\ }\href {https://doi.org/10.22331/q-2024-02-19-1258} {\bibfield  {journal} {\bibinfo  {journal} {Quantum}\ }\textbf {\bibinfo {volume} {8}},\ \bibinfo {pages} {1258} (\bibinfo {year} {2024})}\BibitemShut {NoStop}%
\bibitem [{\citenamefont {Ding}\ \emph {et~al.}(2020)\citenamefont {Ding}, \citenamefont {Wu}, \citenamefont {Holmes}, \citenamefont {Wiseth}, \citenamefont {Franklin}, \citenamefont {Martonosi},\ and\ \citenamefont {Chong}}]{10.1109/ISCA45697.2020.00054}%
  \BibitemOpen
  \bibfield  {author} {\bibinfo {author} {\bibfnamefont {Y.}~\bibnamefont {Ding}}, \bibinfo {author} {\bibfnamefont {X.-C.}\ \bibnamefont {Wu}}, \bibinfo {author} {\bibfnamefont {A.}~\bibnamefont {Holmes}}, \bibinfo {author} {\bibfnamefont {A.}~\bibnamefont {Wiseth}}, \bibinfo {author} {\bibfnamefont {D.}~\bibnamefont {Franklin}}, \bibinfo {author} {\bibfnamefont {M.}~\bibnamefont {Martonosi}},\ and\ \bibinfo {author} {\bibfnamefont {F.~T.}\ \bibnamefont {Chong}},\ }\bibfield  {title} {\bibinfo {title} {Square: strategic quantum ancilla reuse for modular quantum programs via cost-effective uncomputation},\ }in\ \href {https://doi.org/10.1109/ISCA45697.2020.00054} {\emph {\bibinfo {booktitle} {Proceedings of the ACM/IEEE 47th Annual International Symposium on Computer Architecture}}},\ \bibinfo {series and number} {ISCA '20}\ (\bibinfo  {publisher} {IEEE Press},\ \bibinfo {year} {2020})\ p.\ \bibinfo {pages} {570–583}\BibitemShut {NoStop}%
\bibitem [{\citenamefont {Goldfriend}\ \emph {et~al.}(2024)\citenamefont {Goldfriend}, \citenamefont {Reichental}, \citenamefont {Naveh}, \citenamefont {Gazit}, \citenamefont {Yoran}, \citenamefont {Alon}, \citenamefont {Ur}, \citenamefont {Lahav}, \citenamefont {Cornfeld}, \citenamefont {Elazari}, \citenamefont {Emanuel}, \citenamefont {Harpaz}, \citenamefont {Michaeli}, \citenamefont {Erez}, \citenamefont {Preminger}, \citenamefont {Shapira}, \citenamefont {Garcell}, \citenamefont {Samimi}, \citenamefont {Kisch}, \citenamefont {Hallel}, \citenamefont {Kishony}, \citenamefont {van Wingerden}, \citenamefont {Rosenbloom}, \citenamefont {Opher}, \citenamefont {Vax}, \citenamefont {Smoler}, \citenamefont {Danzig}, \citenamefont {Schirman}, \citenamefont {Sella}, \citenamefont {Cohen}, \citenamefont {Garfunkel}, \citenamefont {Cohn}, \citenamefont {Rosemarin}, \citenamefont {Hass}, \citenamefont {Jankiewicz}, \citenamefont {Gharra}, \citenamefont {Roth}, \citenamefont {Azar}, \citenamefont {Asban}, \citenamefont
  {Linkov}, \citenamefont {Segman}, \citenamefont {Sahar}, \citenamefont {Davidson}, \citenamefont {Minerbi},\ and\ \citenamefont {Naveh}}]{classiq}%
  \BibitemOpen
  \bibfield  {author} {\bibinfo {author} {\bibfnamefont {T.}~\bibnamefont {Goldfriend}}, \bibinfo {author} {\bibfnamefont {I.}~\bibnamefont {Reichental}}, \bibinfo {author} {\bibfnamefont {A.}~\bibnamefont {Naveh}}, \bibinfo {author} {\bibfnamefont {L.}~\bibnamefont {Gazit}}, \bibinfo {author} {\bibfnamefont {N.}~\bibnamefont {Yoran}}, \bibinfo {author} {\bibfnamefont {R.}~\bibnamefont {Alon}}, \bibinfo {author} {\bibfnamefont {S.}~\bibnamefont {Ur}}, \bibinfo {author} {\bibfnamefont {S.}~\bibnamefont {Lahav}}, \bibinfo {author} {\bibfnamefont {E.}~\bibnamefont {Cornfeld}}, \bibinfo {author} {\bibfnamefont {A.}~\bibnamefont {Elazari}}, \bibinfo {author} {\bibfnamefont {P.}~\bibnamefont {Emanuel}}, \bibinfo {author} {\bibfnamefont {D.}~\bibnamefont {Harpaz}}, \bibinfo {author} {\bibfnamefont {T.}~\bibnamefont {Michaeli}}, \bibinfo {author} {\bibfnamefont {N.}~\bibnamefont {Erez}}, \bibinfo {author} {\bibfnamefont {L.}~\bibnamefont {Preminger}}, \bibinfo {author} {\bibfnamefont {R.}~\bibnamefont {Shapira}},
  \bibinfo {author} {\bibfnamefont {E.~M.}\ \bibnamefont {Garcell}}, \bibinfo {author} {\bibfnamefont {O.}~\bibnamefont {Samimi}}, \bibinfo {author} {\bibfnamefont {S.}~\bibnamefont {Kisch}}, \bibinfo {author} {\bibfnamefont {G.}~\bibnamefont {Hallel}}, \bibinfo {author} {\bibfnamefont {G.}~\bibnamefont {Kishony}}, \bibinfo {author} {\bibfnamefont {V.}~\bibnamefont {van Wingerden}}, \bibinfo {author} {\bibfnamefont {N.~A.}\ \bibnamefont {Rosenbloom}}, \bibinfo {author} {\bibfnamefont {O.}~\bibnamefont {Opher}}, \bibinfo {author} {\bibfnamefont {M.}~\bibnamefont {Vax}}, \bibinfo {author} {\bibfnamefont {A.}~\bibnamefont {Smoler}}, \bibinfo {author} {\bibfnamefont {T.}~\bibnamefont {Danzig}}, \bibinfo {author} {\bibfnamefont {E.}~\bibnamefont {Schirman}}, \bibinfo {author} {\bibfnamefont {G.}~\bibnamefont {Sella}}, \bibinfo {author} {\bibfnamefont {R.}~\bibnamefont {Cohen}}, \bibinfo {author} {\bibfnamefont {R.}~\bibnamefont {Garfunkel}}, \bibinfo {author} {\bibfnamefont {T.}~\bibnamefont {Cohn}}, \bibinfo
  {author} {\bibfnamefont {H.}~\bibnamefont {Rosemarin}}, \bibinfo {author} {\bibfnamefont {R.}~\bibnamefont {Hass}}, \bibinfo {author} {\bibfnamefont {K.}~\bibnamefont {Jankiewicz}}, \bibinfo {author} {\bibfnamefont {K.}~\bibnamefont {Gharra}}, \bibinfo {author} {\bibfnamefont {O.}~\bibnamefont {Roth}}, \bibinfo {author} {\bibfnamefont {B.}~\bibnamefont {Azar}}, \bibinfo {author} {\bibfnamefont {S.}~\bibnamefont {Asban}}, \bibinfo {author} {\bibfnamefont {N.}~\bibnamefont {Linkov}}, \bibinfo {author} {\bibfnamefont {D.}~\bibnamefont {Segman}}, \bibinfo {author} {\bibfnamefont {O.}~\bibnamefont {Sahar}}, \bibinfo {author} {\bibfnamefont {N.}~\bibnamefont {Davidson}}, \bibinfo {author} {\bibfnamefont {N.}~\bibnamefont {Minerbi}},\ and\ \bibinfo {author} {\bibfnamefont {Y.}~\bibnamefont {Naveh}},\ }\href {https://arxiv.org/abs/2412.07372} {\bibinfo {title} {Design and synthesis of scalable quantum programs}} (\bibinfo {year} {2024}),\ \Eprint {https://arxiv.org/abs/2412.07372} {arXiv:2412.07372 [quant-ph]}
  \BibitemShut {NoStop}%
\bibitem [{\citenamefont {Cross}\ \emph {et~al.}(2017)\citenamefont {Cross}, \citenamefont {Bishop}, \citenamefont {Smolin},\ and\ \citenamefont {Gambetta}}]{openqasm}%
  \BibitemOpen
  \bibfield  {author} {\bibinfo {author} {\bibfnamefont {A.~W.}\ \bibnamefont {Cross}}, \bibinfo {author} {\bibfnamefont {L.~S.}\ \bibnamefont {Bishop}}, \bibinfo {author} {\bibfnamefont {J.~A.}\ \bibnamefont {Smolin}},\ and\ \bibinfo {author} {\bibfnamefont {J.~M.}\ \bibnamefont {Gambetta}},\ }\href {https://arxiv.org/abs/1707.03429} {\bibinfo {title} {Open quantum assembly language}} (\bibinfo {year} {2017}),\ \Eprint {https://arxiv.org/abs/1707.03429} {arXiv:1707.03429 [quant-ph]} \BibitemShut {NoStop}%
\bibitem [{\citenamefont {{QIR Alliance}}(2021)}]{qir}%
  \BibitemOpen
  \bibfield  {author} {\bibinfo {author} {\bibnamefont {{QIR Alliance}}},\ }\href {https://github.com/qir-alliance/qir-spec} {\emph {\bibinfo {title} {{QIR Specification}}}} (\bibinfo {year} {2021}),\ \bibinfo {note} {see \url{https://qir-alliance.org}}\BibitemShut {NoStop}%
\bibitem [{\citenamefont {Bichsel}\ \emph {et~al.}(2020)\citenamefont {Bichsel}, \citenamefont {Baader}, \citenamefont {Gehr},\ and\ \citenamefont {Vechev}}]{bichsel20}%
  \BibitemOpen
  \bibfield  {author} {\bibinfo {author} {\bibfnamefont {B.}~\bibnamefont {Bichsel}}, \bibinfo {author} {\bibfnamefont {M.}~\bibnamefont {Baader}}, \bibinfo {author} {\bibfnamefont {T.}~\bibnamefont {Gehr}},\ and\ \bibinfo {author} {\bibfnamefont {M.}~\bibnamefont {Vechev}},\ }\bibfield  {title} {\bibinfo {title} {Silq: a high-level quantum language with safe uncomputation and intuitive semantics},\ }in\ \href {https://doi.org/10.1145/3385412.3386007} {\emph {\bibinfo {booktitle} {Proceedings of the 41st ACM SIGPLAN Conference on Programming Language Design and Implementation}}},\ \bibinfo {series and number} {PLDI 2020}\ (\bibinfo  {publisher} {Association for Computing Machinery},\ \bibinfo {address} {New York, NY, USA},\ \bibinfo {year} {2020})\ p.\ \bibinfo {pages} {286–300}\BibitemShut {NoStop}%
\bibitem [{\citenamefont {Vax}\ \emph {et~al.}(2025)\citenamefont {Vax}, \citenamefont {Emanuel}, \citenamefont {Cornfeld}, \citenamefont {Reichental}, \citenamefont {Opher}, \citenamefont {Roth}, \citenamefont {Michaeli}, \citenamefont {Preminger}, \citenamefont {Gazit}, \citenamefont {Naveh},\ and\ \citenamefont {Naveh}}]{vax2025qmodexpressivehighlevelquantum}%
  \BibitemOpen
  \bibfield  {author} {\bibinfo {author} {\bibfnamefont {M.}~\bibnamefont {Vax}}, \bibinfo {author} {\bibfnamefont {P.}~\bibnamefont {Emanuel}}, \bibinfo {author} {\bibfnamefont {E.}~\bibnamefont {Cornfeld}}, \bibinfo {author} {\bibfnamefont {I.}~\bibnamefont {Reichental}}, \bibinfo {author} {\bibfnamefont {O.}~\bibnamefont {Opher}}, \bibinfo {author} {\bibfnamefont {O.}~\bibnamefont {Roth}}, \bibinfo {author} {\bibfnamefont {T.}~\bibnamefont {Michaeli}}, \bibinfo {author} {\bibfnamefont {L.}~\bibnamefont {Preminger}}, \bibinfo {author} {\bibfnamefont {L.}~\bibnamefont {Gazit}}, \bibinfo {author} {\bibfnamefont {A.}~\bibnamefont {Naveh}},\ and\ \bibinfo {author} {\bibfnamefont {Y.}~\bibnamefont {Naveh}},\ }\href {https://arxiv.org/abs/2502.19368} {\bibinfo {title} {Qmod: Expressive high-level quantum modeling}} (\bibinfo {year} {2025}),\ \Eprint {https://arxiv.org/abs/2502.19368} {arXiv:2502.19368 [quant-ph]} \BibitemShut {NoStop}%
\bibitem [{\citenamefont {Javadi-Abhari}\ \emph {et~al.}(2024)\citenamefont {Javadi-Abhari}, \citenamefont {Treinish}, \citenamefont {Krsulich}, \citenamefont {Wood}, \citenamefont {Lishman}, \citenamefont {Gacon}, \citenamefont {Martiel}, \citenamefont {Nation}, \citenamefont {Bishop}, \citenamefont {Cross}, \citenamefont {Johnson},\ and\ \citenamefont {Gambetta}}]{qiskit}%
  \BibitemOpen
  \bibfield  {author} {\bibinfo {author} {\bibfnamefont {A.}~\bibnamefont {Javadi-Abhari}}, \bibinfo {author} {\bibfnamefont {M.}~\bibnamefont {Treinish}}, \bibinfo {author} {\bibfnamefont {K.}~\bibnamefont {Krsulich}}, \bibinfo {author} {\bibfnamefont {C.~J.}\ \bibnamefont {Wood}}, \bibinfo {author} {\bibfnamefont {J.}~\bibnamefont {Lishman}}, \bibinfo {author} {\bibfnamefont {J.}~\bibnamefont {Gacon}}, \bibinfo {author} {\bibfnamefont {S.}~\bibnamefont {Martiel}}, \bibinfo {author} {\bibfnamefont {P.~D.}\ \bibnamefont {Nation}}, \bibinfo {author} {\bibfnamefont {L.~S.}\ \bibnamefont {Bishop}}, \bibinfo {author} {\bibfnamefont {A.~W.}\ \bibnamefont {Cross}}, \bibinfo {author} {\bibfnamefont {B.~R.}\ \bibnamefont {Johnson}},\ and\ \bibinfo {author} {\bibfnamefont {J.~M.}\ \bibnamefont {Gambetta}},\ }\href {https://arxiv.org/abs/2405.08810} {\bibinfo {title} {Quantum computing with qiskit}} (\bibinfo {year} {2024}),\ \Eprint {https://arxiv.org/abs/2405.08810} {arXiv:2405.08810 [quant-ph]} \BibitemShut
  {NoStop}%
\bibitem [{\citenamefont {{Microsoft}}()}]{qsharp}%
  \BibitemOpen
  \bibfield  {author} {\bibinfo {author} {\bibnamefont {{Microsoft}}},\ }\href {https://github.com/microsoft/qsharp} {\emph {\bibinfo {title} {{Azure Quantum Development Kit}}}},\ \bibinfo {note} {see \url{https://github.com/microsoft/qsharp}}\BibitemShut {NoStop}%
\bibitem [{\citenamefont {Seidel}\ \emph {et~al.}(2024)\citenamefont {Seidel}, \citenamefont {Bock}, \citenamefont {Zander}, \citenamefont {Petrič}, \citenamefont {Steinmann}, \citenamefont {Tcholtchev},\ and\ \citenamefont {Hauswirth}}]{qrisp}%
  \BibitemOpen
  \bibfield  {author} {\bibinfo {author} {\bibfnamefont {R.}~\bibnamefont {Seidel}}, \bibinfo {author} {\bibfnamefont {S.}~\bibnamefont {Bock}}, \bibinfo {author} {\bibfnamefont {R.}~\bibnamefont {Zander}}, \bibinfo {author} {\bibfnamefont {M.}~\bibnamefont {Petrič}}, \bibinfo {author} {\bibfnamefont {N.}~\bibnamefont {Steinmann}}, \bibinfo {author} {\bibfnamefont {N.}~\bibnamefont {Tcholtchev}},\ and\ \bibinfo {author} {\bibfnamefont {M.}~\bibnamefont {Hauswirth}},\ }\href {https://arxiv.org/abs/2406.14792} {\bibinfo {title} {Qrisp: A framework for compilable high-level programming of gate-based quantum computers}} (\bibinfo {year} {2024}),\ \Eprint {https://arxiv.org/abs/2406.14792} {arXiv:2406.14792 [quant-ph]} \BibitemShut {NoStop}%
\bibitem [{\citenamefont {Sivarajah}\ \emph {et~al.}(2020)\citenamefont {Sivarajah}, \citenamefont {Dilkes}, \citenamefont {Cowtan}, \citenamefont {Simmons}, \citenamefont {Edgington},\ and\ \citenamefont {Duncan}}]{tket}%
  \BibitemOpen
  \bibfield  {author} {\bibinfo {author} {\bibfnamefont {S.}~\bibnamefont {Sivarajah}}, \bibinfo {author} {\bibfnamefont {S.}~\bibnamefont {Dilkes}}, \bibinfo {author} {\bibfnamefont {A.}~\bibnamefont {Cowtan}}, \bibinfo {author} {\bibfnamefont {W.}~\bibnamefont {Simmons}}, \bibinfo {author} {\bibfnamefont {A.}~\bibnamefont {Edgington}},\ and\ \bibinfo {author} {\bibfnamefont {R.}~\bibnamefont {Duncan}},\ }\bibfield  {title} {\bibinfo {title} {t|ket⟩: a retargetable compiler for nisq devices},\ }\href {https://doi.org/10.1088/2058-9565/ab8e92} {\bibfield  {journal} {\bibinfo  {journal} {Quantum Science and Technology}\ }\textbf {\bibinfo {volume} {6}},\ \bibinfo {pages} {014003} (\bibinfo {year} {2020})}\BibitemShut {NoStop}%
\bibitem [{\citenamefont {Kim}\ \emph {et~al.}(2023)\citenamefont {Kim}, \citenamefont {McCaskey}, \citenamefont {Heim}, \citenamefont {Modani}, \citenamefont {Stanwyck},\ and\ \citenamefont {Costa}}]{cuda}%
  \BibitemOpen
  \bibfield  {author} {\bibinfo {author} {\bibfnamefont {J.-S.}\ \bibnamefont {Kim}}, \bibinfo {author} {\bibfnamefont {A.}~\bibnamefont {McCaskey}}, \bibinfo {author} {\bibfnamefont {B.}~\bibnamefont {Heim}}, \bibinfo {author} {\bibfnamefont {M.}~\bibnamefont {Modani}}, \bibinfo {author} {\bibfnamefont {S.}~\bibnamefont {Stanwyck}},\ and\ \bibinfo {author} {\bibfnamefont {T.}~\bibnamefont {Costa}},\ }\bibfield  {title} {\bibinfo {title} {{\relax CUDA Quantum:} the platform for integrated quantum-classical computing},\ }in\ \href {https://doi.org/10.1109/DAC56929.2023.10247886} {\emph {\bibinfo {booktitle} {2023 60th ACM/IEEE Design Automation Conference (DAC)}}}\ (\bibinfo {year} {2023})\ pp.\ \bibinfo {pages} {1--4}\BibitemShut {NoStop}%
\bibitem [{\citenamefont {Bergholm}\ \emph {et~al.}(2022)\citenamefont {Bergholm}, \citenamefont {Izaac}, \citenamefont {Schuld}, \citenamefont {Gogolin}, \citenamefont {Ahmed}, \citenamefont {Ajith}, \citenamefont {Alam}, \citenamefont {Alonso-Linaje}, \citenamefont {AkashNarayanan}, \citenamefont {Asadi}, \citenamefont {Arrazola}, \citenamefont {Azad}, \citenamefont {Banning}, \citenamefont {Blank}, \citenamefont {Bromley}, \citenamefont {Cordier}, \citenamefont {Ceroni}, \citenamefont {Delgado}, \citenamefont {Matteo}, \citenamefont {Dusko}, \citenamefont {Garg}, \citenamefont {Guala}, \citenamefont {Hayes}, \citenamefont {Hill}, \citenamefont {Ijaz}, \citenamefont {Isacsson}, \citenamefont {Ittah}, \citenamefont {Jahangiri}, \citenamefont {Jain}, \citenamefont {Jiang}, \citenamefont {Khandelwal}, \citenamefont {Kottmann}, \citenamefont {Lang}, \citenamefont {Lee}, \citenamefont {Loke}, \citenamefont {Lowe}, \citenamefont {McKiernan}, \citenamefont {Meyer}, \citenamefont {Montañez-Barrera}, \citenamefont
  {Moyard}, \citenamefont {Niu}, \citenamefont {O'Riordan}, \citenamefont {Oud}, \citenamefont {Panigrahi}, \citenamefont {Park}, \citenamefont {Polatajko}, \citenamefont {Quesada}, \citenamefont {Roberts}, \citenamefont {Sá}, \citenamefont {Schoch}, \citenamefont {Shi}, \citenamefont {Shu}, \citenamefont {Sim}, \citenamefont {Singh}, \citenamefont {Strandberg}, \citenamefont {Soni}, \citenamefont {Száva}, \citenamefont {Thabet}, \citenamefont {Vargas-Hernández}, \citenamefont {Vincent}, \citenamefont {Vitucci}, \citenamefont {Weber}, \citenamefont {Wierichs}, \citenamefont {Wiersema}, \citenamefont {Willmann}, \citenamefont {Wong}, \citenamefont {Zhang},\ and\ \citenamefont {Killoran}}]{pennylane}%
  \BibitemOpen
  \bibfield  {author} {\bibinfo {author} {\bibfnamefont {V.}~\bibnamefont {Bergholm}}, \bibinfo {author} {\bibfnamefont {J.}~\bibnamefont {Izaac}}, \bibinfo {author} {\bibfnamefont {M.}~\bibnamefont {Schuld}}, \bibinfo {author} {\bibfnamefont {C.}~\bibnamefont {Gogolin}}, \bibinfo {author} {\bibfnamefont {S.}~\bibnamefont {Ahmed}}, \bibinfo {author} {\bibfnamefont {V.}~\bibnamefont {Ajith}}, \bibinfo {author} {\bibfnamefont {M.~S.}\ \bibnamefont {Alam}}, \bibinfo {author} {\bibfnamefont {G.}~\bibnamefont {Alonso-Linaje}}, \bibinfo {author} {\bibfnamefont {B.}~\bibnamefont {AkashNarayanan}}, \bibinfo {author} {\bibfnamefont {A.}~\bibnamefont {Asadi}}, \bibinfo {author} {\bibfnamefont {J.~M.}\ \bibnamefont {Arrazola}}, \bibinfo {author} {\bibfnamefont {U.}~\bibnamefont {Azad}}, \bibinfo {author} {\bibfnamefont {S.}~\bibnamefont {Banning}}, \bibinfo {author} {\bibfnamefont {C.}~\bibnamefont {Blank}}, \bibinfo {author} {\bibfnamefont {T.~R.}\ \bibnamefont {Bromley}}, \bibinfo {author} {\bibfnamefont {B.~A.}\
  \bibnamefont {Cordier}}, \bibinfo {author} {\bibfnamefont {J.}~\bibnamefont {Ceroni}}, \bibinfo {author} {\bibfnamefont {A.}~\bibnamefont {Delgado}}, \bibinfo {author} {\bibfnamefont {O.~D.}\ \bibnamefont {Matteo}}, \bibinfo {author} {\bibfnamefont {A.}~\bibnamefont {Dusko}}, \bibinfo {author} {\bibfnamefont {T.}~\bibnamefont {Garg}}, \bibinfo {author} {\bibfnamefont {D.}~\bibnamefont {Guala}}, \bibinfo {author} {\bibfnamefont {A.}~\bibnamefont {Hayes}}, \bibinfo {author} {\bibfnamefont {R.}~\bibnamefont {Hill}}, \bibinfo {author} {\bibfnamefont {A.}~\bibnamefont {Ijaz}}, \bibinfo {author} {\bibfnamefont {T.}~\bibnamefont {Isacsson}}, \bibinfo {author} {\bibfnamefont {D.}~\bibnamefont {Ittah}}, \bibinfo {author} {\bibfnamefont {S.}~\bibnamefont {Jahangiri}}, \bibinfo {author} {\bibfnamefont {P.}~\bibnamefont {Jain}}, \bibinfo {author} {\bibfnamefont {E.}~\bibnamefont {Jiang}}, \bibinfo {author} {\bibfnamefont {A.}~\bibnamefont {Khandelwal}}, \bibinfo {author} {\bibfnamefont {K.}~\bibnamefont {Kottmann}},
  \bibinfo {author} {\bibfnamefont {R.~A.}\ \bibnamefont {Lang}}, \bibinfo {author} {\bibfnamefont {C.}~\bibnamefont {Lee}}, \bibinfo {author} {\bibfnamefont {T.}~\bibnamefont {Loke}}, \bibinfo {author} {\bibfnamefont {A.}~\bibnamefont {Lowe}}, \bibinfo {author} {\bibfnamefont {K.}~\bibnamefont {McKiernan}}, \bibinfo {author} {\bibfnamefont {J.~J.}\ \bibnamefont {Meyer}}, \bibinfo {author} {\bibfnamefont {J.~A.}\ \bibnamefont {Montañez-Barrera}}, \bibinfo {author} {\bibfnamefont {R.}~\bibnamefont {Moyard}}, \bibinfo {author} {\bibfnamefont {Z.}~\bibnamefont {Niu}}, \bibinfo {author} {\bibfnamefont {L.~J.}\ \bibnamefont {O'Riordan}}, \bibinfo {author} {\bibfnamefont {S.}~\bibnamefont {Oud}}, \bibinfo {author} {\bibfnamefont {A.}~\bibnamefont {Panigrahi}}, \bibinfo {author} {\bibfnamefont {C.-Y.}\ \bibnamefont {Park}}, \bibinfo {author} {\bibfnamefont {D.}~\bibnamefont {Polatajko}}, \bibinfo {author} {\bibfnamefont {N.}~\bibnamefont {Quesada}}, \bibinfo {author} {\bibfnamefont {C.}~\bibnamefont {Roberts}},
  \bibinfo {author} {\bibfnamefont {N.}~\bibnamefont {Sá}}, \bibinfo {author} {\bibfnamefont {I.}~\bibnamefont {Schoch}}, \bibinfo {author} {\bibfnamefont {B.}~\bibnamefont {Shi}}, \bibinfo {author} {\bibfnamefont {S.}~\bibnamefont {Shu}}, \bibinfo {author} {\bibfnamefont {S.}~\bibnamefont {Sim}}, \bibinfo {author} {\bibfnamefont {A.}~\bibnamefont {Singh}}, \bibinfo {author} {\bibfnamefont {I.}~\bibnamefont {Strandberg}}, \bibinfo {author} {\bibfnamefont {J.}~\bibnamefont {Soni}}, \bibinfo {author} {\bibfnamefont {A.}~\bibnamefont {Száva}}, \bibinfo {author} {\bibfnamefont {S.}~\bibnamefont {Thabet}}, \bibinfo {author} {\bibfnamefont {R.~A.}\ \bibnamefont {Vargas-Hernández}}, \bibinfo {author} {\bibfnamefont {T.}~\bibnamefont {Vincent}}, \bibinfo {author} {\bibfnamefont {N.}~\bibnamefont {Vitucci}}, \bibinfo {author} {\bibfnamefont {M.}~\bibnamefont {Weber}}, \bibinfo {author} {\bibfnamefont {D.}~\bibnamefont {Wierichs}}, \bibinfo {author} {\bibfnamefont {R.}~\bibnamefont {Wiersema}}, \bibinfo {author}
  {\bibfnamefont {M.}~\bibnamefont {Willmann}}, \bibinfo {author} {\bibfnamefont {V.}~\bibnamefont {Wong}}, \bibinfo {author} {\bibfnamefont {S.}~\bibnamefont {Zhang}},\ and\ \bibinfo {author} {\bibfnamefont {N.}~\bibnamefont {Killoran}},\ }\href {https://arxiv.org/abs/1811.04968} {\bibinfo {title} {Pennylane: Automatic differentiation of hybrid quantum-classical computations}} (\bibinfo {year} {2022}),\ \Eprint {https://arxiv.org/abs/1811.04968} {arXiv:1811.04968 [quant-ph]} \BibitemShut {NoStop}%
\bibitem [{\citenamefont {{\relax Cirq Developers}}(2024)}]{cirq}%
  \BibitemOpen
  \bibfield  {author} {\bibinfo {author} {\bibnamefont {{\relax Cirq Developers}}},\ }\href {https://doi.org/10.5281/zenodo.11398048} {\bibinfo {title} {Cirq}} (\bibinfo {year} {2024})\BibitemShut {NoStop}%
\bibitem [{\citenamefont {Yuan}\ and\ \citenamefont {Carbin}(2022)}]{tower}%
  \BibitemOpen
  \bibfield  {author} {\bibinfo {author} {\bibfnamefont {C.}~\bibnamefont {Yuan}}\ and\ \bibinfo {author} {\bibfnamefont {M.}~\bibnamefont {Carbin}},\ }\bibfield  {title} {\bibinfo {title} {Tower: data structures in quantum superposition},\ }\bibfield  {journal} {\bibinfo  {journal} {Proc. ACM Program. Lang.}\ }\textbf {\bibinfo {volume} {6}},\ \href {https://doi.org/10.1145/3563297} {10.1145/3563297} (\bibinfo {year} {2022})\BibitemShut {NoStop}%
\bibitem [{\citenamefont {Barenco}\ \emph {et~al.}(1995)\citenamefont {Barenco}, \citenamefont {Bennett}, \citenamefont {Cleve}, \citenamefont {DiVincenzo}, \citenamefont {Margolus}, \citenamefont {Shor}, \citenamefont {Sleator}, \citenamefont {Smolin},\ and\ \citenamefont {Weinfurter}}]{barenco95}%
  \BibitemOpen
  \bibfield  {author} {\bibinfo {author} {\bibfnamefont {A.}~\bibnamefont {Barenco}}, \bibinfo {author} {\bibfnamefont {C.~H.}\ \bibnamefont {Bennett}}, \bibinfo {author} {\bibfnamefont {R.}~\bibnamefont {Cleve}}, \bibinfo {author} {\bibfnamefont {D.~P.}\ \bibnamefont {DiVincenzo}}, \bibinfo {author} {\bibfnamefont {N.}~\bibnamefont {Margolus}}, \bibinfo {author} {\bibfnamefont {P.}~\bibnamefont {Shor}}, \bibinfo {author} {\bibfnamefont {T.}~\bibnamefont {Sleator}}, \bibinfo {author} {\bibfnamefont {J.~A.}\ \bibnamefont {Smolin}},\ and\ \bibinfo {author} {\bibfnamefont {H.}~\bibnamefont {Weinfurter}},\ }\bibfield  {title} {\bibinfo {title} {Elementary gates for quantum computation},\ }\href {https://doi.org/10.1103/physreva.52.3457} {\bibfield  {journal} {\bibinfo  {journal} {Physical Review A}\ }\textbf {\bibinfo {volume} {52}},\ \bibinfo {pages} {3457–3467} (\bibinfo {year} {1995})}\BibitemShut {NoStop}%
\bibitem [{\citenamefont {Maslov}(2016)}]{PhysRevA.93.022311}%
  \BibitemOpen
  \bibfield  {author} {\bibinfo {author} {\bibfnamefont {D.}~\bibnamefont {Maslov}},\ }\bibfield  {title} {\bibinfo {title} {Advantages of using relative-phase toffoli gates with an application to multiple control toffoli optimization},\ }\href {https://doi.org/10.1103/PhysRevA.93.022311} {\bibfield  {journal} {\bibinfo  {journal} {Phys. Rev. A}\ }\textbf {\bibinfo {volume} {93}},\ \bibinfo {pages} {022311} (\bibinfo {year} {2016})}\BibitemShut {NoStop}%
\bibitem [{\citenamefont {Huang}\ and\ \citenamefont {Palsberg}(2024)}]{10.1145/3656436}%
  \BibitemOpen
  \bibfield  {author} {\bibinfo {author} {\bibfnamefont {K.}~\bibnamefont {Huang}}\ and\ \bibinfo {author} {\bibfnamefont {J.}~\bibnamefont {Palsberg}},\ }\bibfield  {title} {\bibinfo {title} {Compiling conditional quantum gates without using helper qubits},\ }\bibfield  {journal} {\bibinfo  {journal} {Proc. ACM Program. Lang.}\ }\textbf {\bibinfo {volume} {8}},\ \href {https://doi.org/10.1145/3656436} {10.1145/3656436} (\bibinfo {year} {2024})\BibitemShut {NoStop}%
\bibitem [{\citenamefont {Gidney}(2015)}]{Gidney2015}%
  \BibitemOpen
  \bibfield  {author} {\bibinfo {author} {\bibfnamefont {C.}~\bibnamefont {Gidney}},\ }\href {https://algassert.com/circuits/2015/06/05/Constructing-Large-Controlled-Nots.html} {\bibinfo {title} {Constructing large controlled-nots}} (\bibinfo {year} {2015}),\ \bibinfo {note} {see \href{https://algassert.com/circuits/2015/06/05/Constructing-Large-Controlled-Nots.html}{\texttt{https://algassert.com/circuits\hspace{0pt}/2015/06/05/Constructing-Large-Controlled-Nots.html}}}\BibitemShut {NoStop}%
\bibitem [{\citenamefont {Kahn}(1962)}]{kahn62}%
  \BibitemOpen
  \bibfield  {author} {\bibinfo {author} {\bibfnamefont {A.~B.}\ \bibnamefont {Kahn}},\ }\bibfield  {title} {\bibinfo {title} {Topological sorting of large networks},\ }\href {https://doi.org/10.1145/368996.369025} {\bibfield  {journal} {\bibinfo  {journal} {Commun. ACM}\ }\textbf {\bibinfo {volume} {5}},\ \bibinfo {pages} {558–562} (\bibinfo {year} {1962})}\BibitemShut {NoStop}%
\bibitem [{\citenamefont {Purdom}(1970)}]{purdom_transitive_1970}%
  \BibitemOpen
  \bibfield  {author} {\bibinfo {author} {\bibfnamefont {P.}~\bibnamefont {Purdom}},\ }\bibfield  {title} {\bibinfo {title} {A transitive closure algorithm},\ }\href {https://doi.org/10.1007/BF01940892} {\bibfield  {journal} {\bibinfo  {journal} {BIT Numerical Mathematics}\ }\textbf {\bibinfo {volume} {10}},\ \bibinfo {pages} {76} (\bibinfo {year} {1970})}\BibitemShut {NoStop}%
\bibitem [{\citenamefont {Iten}\ \emph {et~al.}(2022)\citenamefont {Iten}, \citenamefont {Moyard}, \citenamefont {Metger}, \citenamefont {Sutter},\ and\ \citenamefont {Woerner}}]{iten22}%
  \BibitemOpen
  \bibfield  {author} {\bibinfo {author} {\bibfnamefont {R.}~\bibnamefont {Iten}}, \bibinfo {author} {\bibfnamefont {R.}~\bibnamefont {Moyard}}, \bibinfo {author} {\bibfnamefont {T.}~\bibnamefont {Metger}}, \bibinfo {author} {\bibfnamefont {D.}~\bibnamefont {Sutter}},\ and\ \bibinfo {author} {\bibfnamefont {S.}~\bibnamefont {Woerner}},\ }\bibfield  {title} {\bibinfo {title} {Exact and practical pattern matching for quantum circuit optimization},\ }\href {https://doi.org/10.1145/3498325} {\bibfield  {journal} {\bibinfo  {journal} {ACM Transactions on Quantum Computing}\ }\textbf {\bibinfo {volume} {3}},\ \bibinfo {pages} {1–41} (\bibinfo {year} {2022})}\BibitemShut {NoStop}%
\bibitem [{\citenamefont {Nam}\ \emph {et~al.}(2018)\citenamefont {Nam}, \citenamefont {Ross}, \citenamefont {Su}, \citenamefont {Childs},\ and\ \citenamefont {Maslov}}]{nam18}%
  \BibitemOpen
  \bibfield  {author} {\bibinfo {author} {\bibfnamefont {Y.}~\bibnamefont {Nam}}, \bibinfo {author} {\bibfnamefont {N.~J.}\ \bibnamefont {Ross}}, \bibinfo {author} {\bibfnamefont {Y.}~\bibnamefont {Su}}, \bibinfo {author} {\bibfnamefont {A.~M.}\ \bibnamefont {Childs}},\ and\ \bibinfo {author} {\bibfnamefont {D.}~\bibnamefont {Maslov}},\ }\bibfield  {title} {\bibinfo {title} {Automated optimization of large quantum circuits with continuous parameters},\ }\href {https://doi.org/10.1038/s41534-018-0072-4} {\bibfield  {journal} {\bibinfo  {journal} {npj Quantum Information}\ }\textbf {\bibinfo {volume} {4}},\ \bibinfo {pages} {23} (\bibinfo {year} {2018})}\BibitemShut {NoStop}%
\bibitem [{\citenamefont {Yuan}\ and\ \citenamefont {Carbin}(2024)}]{10.1145/3656397}%
  \BibitemOpen
  \bibfield  {author} {\bibinfo {author} {\bibfnamefont {C.}~\bibnamefont {Yuan}}\ and\ \bibinfo {author} {\bibfnamefont {M.}~\bibnamefont {Carbin}},\ }\bibfield  {title} {\bibinfo {title} {The t-complexity costs of error correction for control flow in quantum computation},\ }\bibfield  {journal} {\bibinfo  {journal} {Proc. ACM Program. Lang.}\ }\textbf {\bibinfo {volume} {8}},\ \href {https://doi.org/10.1145/3656397} {10.1145/3656397} (\bibinfo {year} {2024})\BibitemShut {NoStop}%
\bibitem [{\citenamefont {Seidel}(2024)}]{seidel24}%
  \BibitemOpen
  \bibfield  {author} {\bibinfo {author} {\bibfnamefont {R.}~\bibnamefont {Seidel}},\ }\bibfield  {title} {\bibinfo {title} {Automatic quantum function parallelization and memory management in qrisp},\ }in\ \href {https://quantum-compilers.github.io/iwqc2024/papers/IWQC2024_paper_16.pdf} {\emph {\bibinfo {booktitle} {Proceedings of the 6th International Workshop on Quantum Compilation}}}\ (\bibinfo {year} {2024})\ \bibinfo {note} {pDF available at \href{https://quantum-compilers.github.io/iwqc2024/papers/IWQC2024_paper_16.pdf}{\texttt{https://quantum-compilers.github.io\hspace{0pt}/iwqc2024/papers/IWQC2024\_paper\_16.pdf}}}\BibitemShut {NoStop}%
\end{thebibliography}%

\end{document}